\documentclass[lettersize,journal]{IEEEtran}
\usepackage{amsmath,amsfonts}
\usepackage{algorithmic}
\usepackage{algorithm}
\usepackage{array}
\usepackage[caption=false,font=scriptsize,labelfont=sf]{subfig}
\usepackage{textcomp}
\usepackage{stfloats}
\usepackage{url}
\usepackage{verbatim}
\usepackage{graphicx}
\usepackage{cite}
\usepackage{bm}
\usepackage{accents}
\usepackage{array}
\usepackage{multicol}
\usepackage{multirow}
\usepackage{eqparbox}
\usepackage{enumerate}
\usepackage{booktabs}
\usepackage{soul}
\usepackage{xcolor}
\usepackage{nccmath}
\usepackage{makecell}

\newcolumntype{M}[1]{>{\centering\arraybackslash}m{#1}}
\newcolumntype{P}[1]{>{\centering\arraybackslash}p{#1}}

\begin{document}
\title{Temporal-Aware Deep Reinforcement Learning for Energy Storage Bidding in Energy and Contingency Reserve Markets}

\author{Jinhao Li,
Changlong Wang,
Yanru Zhang,~\IEEEmembership{Member~IEEE,} and Hao Wang,~\IEEEmembership{Member~IEEE}
\thanks{This work was supported in part by the Australian Research Council (ARC) Discovery Early Career Researcher Award (DECRA) under Grant DE230100046. (Corresponding author: Hao Wang.)}
\thanks{J. Li and H. Wang are with the Department of Data Science and AI, Faculty of IT and Monash Energy Institute, Monash University, Melbourne, VIC 3800, Australia (e-mails: \{jinhao.li, hao.wang2\}@monash.edu).}
\thanks{C. Wang is with the Department of Civil Engineering, Monash University, Melbourne, VIC 3800, Australia  (e-mail: chang.wang@monash.edu).}
\thanks{Y. Zhang is with the School of Computer Science and Engineering, University of Electronic Science and Technology of China, Chengdu, China, 611731 (e-mail: yanruzhang@uestc.edu.cn).}
}



\maketitle
\thispagestyle{empty}

\begin{abstract}
The battery energy storage system (BESS) has immense potential for enhancing grid reliability and security through its participation in the electricity market. BESS often seeks various revenue streams by taking part in multiple markets to unlock its full potential, but effective algorithms for joint-market participation under price uncertainties are insufficiently explored in the existing research. To bridge this gap, we develop a novel BESS joint bidding strategy that utilizes deep reinforcement learning (DRL) to bid in the spot and contingency frequency control ancillary services (FCAS) markets. Our approach leverages a transformer-based temporal feature extractor to effectively respond to price fluctuations in seven markets simultaneously and helps DRL learn the best BESS bidding strategy in joint-market participation. Additionally, unlike conventional ``black-box'' DRL model, our approach is more interpretable and provides valuable insights into the temporal bidding behavior of BESS in the dynamic electricity market.
We validate our method using realistic market prices from the Australian National Electricity Market. The results show that our strategy outperforms benchmarks, including both optimization-based and other DRL-based strategies, by substantial margins. Our findings further suggest that effective temporal-aware bidding can significantly increase profits in the spot and contingency FCAS markets compared to individual market participation.
\end{abstract}


\section{Introduction} \label{sec:intro}
As the world strives towards achieving net-zero emissions, the adoption of variable renewable energy (VRE) sources, such as wind and solar, has become a crucial component of modern power systems' decarbonization efforts~\cite{IPCC2022}. However, the inherently stochastic nature of VRE presents challenges in maintaining power system reliability and security~\cite{sinsel2020}. System reliability refers to the ability of a power system to meet consumer demand at all times, while system security refers to the ability of the system to operate and remain stable in the event of a contingency. In response to the integration challenges posed by VRE, the deployment of grid-scale and community-scale battery energy storage systems (BESS) has increased in recent years. For example, the world's first grid-scale lithium-ion BESS was installed in Australia in 2017, and after five years of successful operation, its role in the modern power system has become increasingly apparent to policymakers. As a result, policymakers have taken proactive measures to promote the deployment of BESS. Notably, the Victorian Government in Australia has introduced energy storage targets aiming to achieve a total capacity of $2.6$GW by 2030 and $6.3$GW by 2035 within the state~\cite{VictoriaGov}.

A BESS can participate in the Australian National Electricity Market (NEM) and generates revenue through two streams, as it can do in other market-based electricity systems. Firstly, the BESS can help maintain system reliability by balancing the mismatch between time-varying generation (with renewables) and demand~\cite{aemo2019}. Such mismatches lead to price fluctuations in the real-time wholesale spot market, creating economic incentives for the BESS to participate in the spot market for energy arbitrage, i.e., buy low and sell high. Secondly, the BESS can provide grid services to enhance system security and stability in the frequency control ancillary services (FCAS) market~\cite{aemo2015}. The provision of these system services was historically a by-product of large synchronous generators in the market, e.g., coal-fired power plants. With coal generators retiring and exiting the market, VRE is growing fast as the main source of supply. As a result, market operators are increasingly compelled to intervene and seek system services from new market participants, particularly from grid-scale BESS~\cite{aemo2021}, highlighting the crucial role of the BESS.

Given the multiple prospective revenue streams that the BESS is exposed to, strategic participation in multiple markets is essential to unlocking the full potential of the BESS in supporting the transitioning grid while maximizing its economic returns. For the BESS, such strategic participation demands the optimal scheduling in spot and FCAS markets simultaneously, commonly referred to as joint bidding. However, this presents notable challenges arising from the highly volatile nature of market prices (as the exogenous uncertainty) and coupling resource constraint (as the endogenous challenge due to limited BESS capacity shared by multiple markets over time). Given these complexities, the development of an effective framework for joint bidding of the BESS across multiple markets is of great value to both the BESS owners and the power system.

Previous research has examined joint bidding using optimization-based approaches. The underlying real-time bidding strategies are derived mostly through stochastic optimization~\cite{abdulla2018,krishnamurthy2018}, whose performance is highly dependent on accurate energy price modeling and forecasting. Predicting energy prices, however, is notoriously difficult since the spot and FCAS markets are highly volatile~\cite{weron2014}, and the price drivers are remarkably complex. Alternatively, deep reinforcement learning (DRL)-based methods~\cite{wang2018,bui2020,cao2020,xu2019,wei2022,huang2020,anwar2022} have drawn increasing attention lately for their data-driven characteristics and interactive learning manner, enabling DRL to dynamically learn the uncertainty of the electricity market without prior knowledge of energy prices or price forecasts.

Existing studies have employed optimization and RL techniques to develop BESS joint-bidding strategies, as discussed in Section~\ref{sec:literature}. Our literature review highlights three research gaps as follows: 1) previous studies tended to overlook the hidden temporal information inside time-varying energy prices. Better decisions could be made by analyzing useful information from the inherent temporal changes of those price signals; 2) ``black-box'' DRL methods lack transparency and interpretability. It is often difficult to trace back how the DRL models understand energy prices and lead to a particular decision in their bidding decision-making mechanism. These methods also cannot provide insights into the charge/discharge behaviors of the BESS when following the proposed bidding strategies, leaving the bidding outcomes less interpretable; 3) joint bidding in multiple markets has not been adequately investigated, particularly for contingency FCAS market. This is noteworthy considering that contingency FCAS constitutes a major revenue source for the BESS in the NEM~\cite{aemo2019_FCAS}.

To bridge above research gaps, we develop a novel temporal-aware DRL-based bidding strategy for the BESS taking part in the spot and contingency FCAS markets simultaneously. Our strategy draws on a transformer-based temporal feature extractor (TTFE) to fully exploit the temporal price spreads of multi-time-series energy prices in both markets for learning a better joint bidding strategy. Specifically, this ``temporal-aware'' capability could better assist the BESS in scheduling charge/discharge for energy arbitrage in the spot market, while concurrently bidding power for contingency FCAS delivery to maximize the overall economic returns. Despite the complexity of joint bidding, our proposed strategy is more interpretable and can shed light on the temporal bidding behaviors of the BESS. The main contributions of our work are summarized as follows.
\begin{itemize}
    \item \textit{BESS Joint-Market Participation:} To meet the increasing need for the provision of frequency services to stabilize the grid, we deploy the BESS to participate in six contingency FCAS sub-markets in addition to the revenue stream through energy arbitrage in the spot market. The facilitation of BESS in joint-market bidding, particularly in the prospective FCAS market, takes advantage of the markets' flexibility and unlocks BESS's economic potential. Our study provides a viable case for BESS's revenue creation in an increasingly complicated electricity market with exogenous uncertainty in prices.
    
    \item \textit{Extracting Temporal Information of Energy Prices:} Unlike previous studies that overlooked the temporal trends of historical energy prices, we develop a novel TTFE with a stacked multi-head attention mechanism to exploit historical multi-market energy prices, extract their underlying temporal information, and provide insights for the BESS to be temporal-aware of and responsive to market volatility for better bidding decisions.

    \item \textit{DRL-based Bidding Strategy:} To address the uncertainty of the real-time markets, we introduce an off-policy DRL algorithm, namely soft actor-critic (SAC), to maximize the overall revenue in the joint bidding problem, modeled as a Markov decision process (MDP). Numerical results demonstrate the effectiveness of our method in creating a significant performance boost and surpassing the optimization-based benchmark by approximately $24\%$.

    \item \textit{Novel Interpretations on ``Black-box DRL'' Models}: Our study offers three novel model-based interpretations of the internal decision-making mechanism of the DRL-based joint bidding strategy. Specifically, the Q-value-based, attention-based, and gradient-based interpretations suggest that capturing historical price fluctuations plays a crucial role in improving bidding performance. The interpretability of our model provides human-understandable insights, increasing trust of using such models in real-world industries.
\end{itemize}

The remainder of this paper is organized as follows. Section \ref{sec:literature} reviews the related work. Section \ref{sec:model} formulates the joint-bidding problem of the BESS in the spot and contingency FCAS markets. Section \ref{sec:method} proposes our temporal-aware DRL-based bidding strategy. Section \ref{sec:exp} presents and discusses simulation results. Section \ref{sec:conclusions} concludes this paper.

\section{Related Work} \label{sec:literature}
Real-time bidding strategies for the BESS have been widely studied using optimization-based methods. For instance, the studies in~\cite{bradbury2014,mcConnell2015,zafirakis2016} approached energy arbitrage as a linear programming problem in the electricity markets of the U.S., Australia, and Europe, respectively. However, these works relied on perfect knowledge of energy prices, which are not available in practice. To address the uncertainty of energy prices, researchers have proposed alternative approaches. Abdulla \textit{et al.}~\cite{abdulla2018} proposed a stochastic dynamic programming approach that used available forecasts to operate a BESS in the wholesale spot market. Similarly, Krishnamurthy \textit{et al.}~\cite{krishnamurthy2018} formulated a stochastic optimization problem that employed price scenario generation to forecast electricity prices. However, both of these studies were dependent on accurate energy price forecasting. Though He \textit{et al.}~\cite{he2023} has designed a real-time bidding strategy for the BESS, they focused on participating in the local electricity market with a community-scale BESS rather than the wholesale market for broader benefits. In addition, 
Chen \textit{et al.}~\cite{chen2021} proposed game-theoretical frameworks to promote energy trading. However, these methods required information from other market participants, making them less applicable in real-world applications.

Recently, bidding strategies using DRL have gained popularity due to their model-free and data-driven characteristics. Several studies, e.g., in \cite{wang2018,wang2023,bui2020,cao2020}, have proposed Q-learning-based bidding strategies for energy arbitrage. These studies utilized various techniques including Q-learning~\cite{wang2018,wang2023}, double deep Q-learning~\cite{bui2020}, and multiple deep Q-learning variants~\cite{cao2020}. Specifically, the works by~\cite{wang2018,bui2020,cao2020} studied real-time bidding in the wholesale electricity market with grid-scale BESSs, while Wang \textit{et al.}~\cite{wang2023} mainly analyzed peer-to-peer energy trading with relatively small-capacity BESSs. The algorithms employed in the above studies discretized the bidding decision space, simplifying the problem at the cost of performance to an extent. In contrast, other works such as~\cite{xu2019,jeong2023} and~\cite{wei2022} employed proximal policy optimization (PPO)~\cite{xu2019,jeong2023} and SAC~\cite{wei2022} algorithms to perform energy arbitrage in the continuous decision space, respectively. Specifically, Jeong \textit{et al.}~\cite{jeong2023} combined renewable generators with BESS in real-time market participation, while the BESS only serves as an onsite backup source to supplement the renewable generation. Moreover, Huang \textit{et al.}~\cite{huang2020} and Anwar \textit{et al.}~\cite{anwar2022} proposed PPO-based joint bidding strategies in the spot and regulation FCAS markets. These approaches aimed to offer a more nuanced approach to bidding in energy markets.

\begin{figure}[!t]
    \centering
    \includegraphics[width=\linewidth]{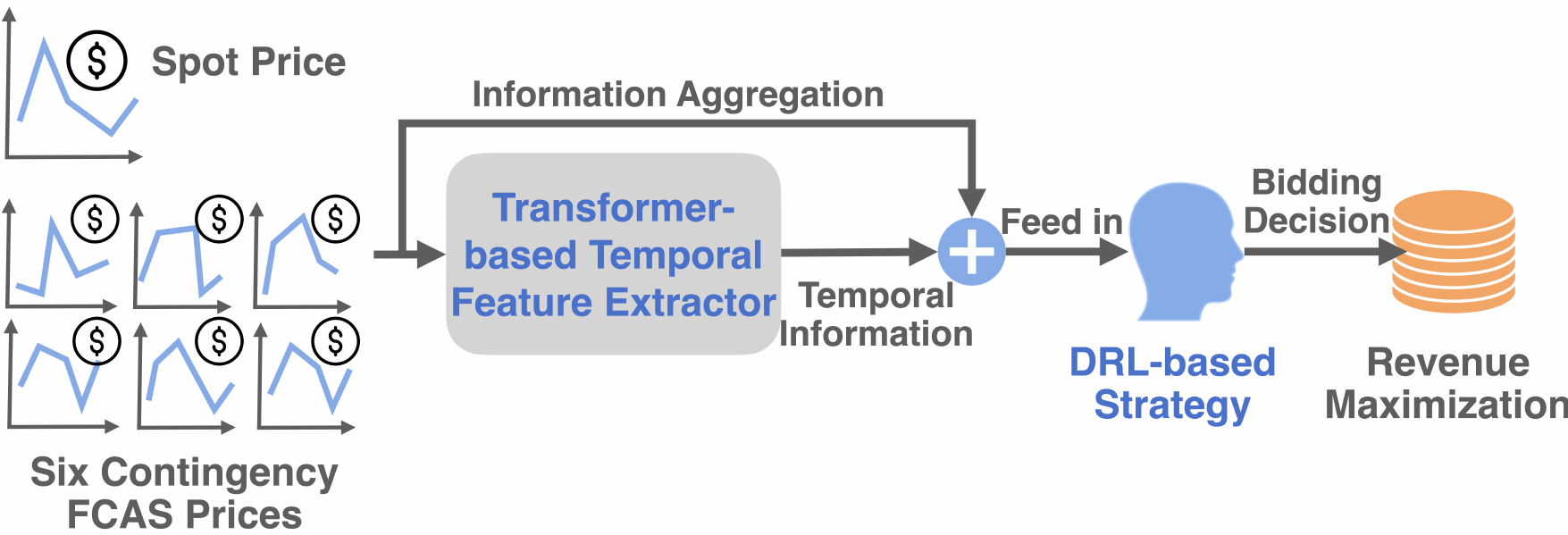}
    \caption{The framework of the TempDRL.}
    \label{fig:framework}
\end{figure}

The literature review suggests that bidding in the contingency FCAS market has received limited attention. To address this research gap and evaluate the potential of joint-market bidding, we present a new approach named ``TempDRL'', enabling the BESS to simultaneously participate in the spot and contingency FCAS markets. Moreover, previous studies tended to neglect the importance of historical price signals and thus did not capture the price trends. While several studies have explored capturing historical temporal information using recurrent neural network structures in the field of robotics~\cite{böhm2023_featureExtraction} and chemical engineering~\cite{liu2023_featureExtraction}, to the best of our knowledge, extracting useful temporal price information from raw price data has been seemingly less discussed. Therefore, to take advantage of historical prices, our approach utilizes a TTFE to effectively extract and use the temporal information from multi-time-series energy prices in both the spot and contingency FCAS markets. This information is then fed into the SAC algorithm to learn a joint bidding strategy that maximizes the overall revenue from both markets. A visual representation of the TempDRL framework is presented in Fig. \ref{fig:framework}.

\section{System Model}\label{sec:model}
In this study, we investigate the participation of a BESS in the NEM as a price-taker, assuming its bids will not affect the market-clearing outcomes, which is reasonable in particular in wholesale markets with high trading volumes. We maximize the potential of the BESS for revenue creation through simultaneous participation in both the spot and contingency FCAS markets. The context of the joint bidding strategy is presented in detail in Section \ref{subsec:model_preliminary_NEM}. Section \ref{subsec:model_revenue_streams} outlines multiple revenue streams of the BESS under various operational conditions. The joint bidding optimization problem is formulated in Section \ref{subsec:model_joint}.

\subsection{The Australian National Electricity Market} \label{subsec:model_preliminary_NEM}

\subsubsection{The Spot Market} \label{subsubsec:model_preliminary_NEM_spot}
As a major part of the NEM, the spot market is a real-time market for trading wholesale electricity between generators and loads, where power supply and demand are balanced instantaneously through a centrally coordinated dispatch process managed by the Australian Energy Market Operator (AEMO)~\cite{aemo2020}. Generators submit bids (price and quantity) every five minutes. AEMO dispatches generators in a least-cost manner by ranking generator bids from low to high to form a bidding stack. The generator bids that fulfill the last power demand in the bidding stack determine the market clearing price, known as the spot price. Generators that bid below or at that price will get dispatched at their offer quantity and get paid at the spot price. The spot price is constantly subject to fluctuations based on the electricity supply-demand mismatch and can increase in response to generation shortages and decrease with redundant generation.

\subsubsection{The Contingency FCAS Market} \label{subsubsec:model_preliminary_NEM_contingencyFCAS}
In the NEM, the FCAS market is established to ensure stable system frequency by procuring reserves in response to increasing penetration of VRE resources~\cite{aemo2015}. The grid-scale BESS can provide two main types of ancillary services: regulation FCAS and contingency FCAS. In this study, we focus on the joint bidding of the BESS in the spot and contingency FCAS markets.

Contingency events, such as power plant failures and transmission network faults, can result in system frequency deviation from the normal operating band (NOB) from $49.95$ to $50.15$Hz. To restore system frequency in these situations, the contingency FCAS market has been established to provide a larger amount of energy for a longer period of time than regulation services~\cite{aemo2015}. This market is divided into six sub-markets, including fast raise (FR), fast lower (FL), slow raise (SR), slow lower (SL), delayed raise (DR), and delayed lower (DL), each requiring different response times of $6$ seconds, $60$ seconds, or $5$-minutes for \textit{fast}, \textit{slow}, and \textit{delayed} sub-markets, respectively. For example, the BESS may bid to discharge in the FR sub-market to arrest a rapid frequency drop within $6$ seconds, or in the SR sub-market to stabilize frequency after a major drop within $60$ seconds, or in the DR sub-market to recover frequency back to the NOB within $5$ minutes.

\subsection{BESS Multi-Market Revenue Streams} \label{subsec:model_revenue_streams}
Energy arbitrage in the spot market and network service provision in the FCAS market are two major revenue streams for the BESS.

\subsubsection{Spot Market} \label{sec:model_revenue_streams_spot}
Real-time power supply-demand mismatches are reflected by price fluctuations in the spot market, creating economic incentives for the BESS to perform energy arbitrage (i.e., buy low and sell high) in the spot market. Given that the BESS cannot simultaneously charge and discharge, we introduce two binary variables $v_t^\text{ch}$ and $v_t^\text{dch}$ to restrict the BESS operation, which can be formulated as
\begin{equation}
    \label{eq:dch_ch_cons}
    v_t^\text{ch} + v_t^\text{dch} \leq 1, \quad v_t^\text{ch},v_t^\text{dch}\in\left\{0,1\right\},
\end{equation}
where the BESS sits idle when these two variables are zero.

We denote the BESS's bid power in the spot market as $p_t^\text{S}$ and let $\rho_t^\text{S}$ be the spot price, $\Delta t$ be the duration of the NEM dispatch interval (i.e., $5$ minutes), $T$ be the total time slots in the operational horizon, and $\eta^\text{ch}, \eta^\text{dch}$ be charging/discharging efficiencies of the BESS. The revenue from the spot market can be formulated as 
\begin{equation}
    \label{eq:revenue_spot}
    R^\text{S} = \Delta t\sum_{t=1}^T \left(v_t^\text{dch}\eta^\text{dch}-v_t^\text{ch}\frac{1}{\eta^\text{ch}}\right)\rho_t^\text{S}p_t^\text{S}.
\end{equation}

\subsubsection{Contingency FCAS Market} \label{subsec:model_revenue_streams_contingencyFCAS}
In addition to energy arbitrage in the spot market, the BESS can deliver contingency frequency services in the six contingency FCAS sub-markets, categorized into raise and lower components across a fast (6 second), slow (60 second), and delayed (5 minute) time frames. According to the associated market rules~\cite{aemo2015}, market participants that bid in the contingency FCAS market get dispatched by the AEMO only in the event of a contingency. Reserved power for service delivery cannot be adopted for other usages, even though the contingency event does not occur. We denote the market clearing prices of the six contingency FCAS sub-markets as $\rho_t^\text{FR}$, $\rho_t^\text{FL}$, $\rho_t^\text{SR}$, $\rho_t^\text{SL}$, $\rho_t^\text{DR}$, and $\rho_t^\text{DL}$ for fast raise, fast lower, slow raise, slow lower, delayed raise, and delayed lower services, respectively~\cite{li2022_workshopPaper}. The BESS's bid power in the fast, slow, and delayed sub-markets are defined as $p_t^\text{fast}$, $p_t^\text{slow}$, and $p_t^\text{delay}$, respectively. Revenue from the contingency FCAS market is formulated as
\begin{align}
\label{eq:revenue_FCAS}
    R^\text{FCAS} &= \Delta t \sum_{t=1}^T   \left[v_t^\text{dch}\eta^\text{dch}\left(\rho_t^\text{FR}p_t^\text{fast}+\rho_t^\text{SR}p_t^\text{slow}+\rho_t^\text{DR}p_t^\text{delay}\right)\right. \nonumber\\
    &\hspace{2em}\left.+v_t^\text{ch}\frac{1}{\eta^\text{ch}}\left(\rho_t^\text{FL}p_t^\text{fast}+\rho_t^\text{SL}p_t^\text{slow}+\rho_t^\text{DL}p_t^\text{delay}\right)\right].
\end{align}

Note that when the BESS chooses to discharge, i.e., $v_t^\text{dch}=1$, the BESS, besides arbitrage in the spot market, is able to deliver frequency-raise services in the FR, SR, and DR sub-markets but cannot provide services that involve charging. On the contrary, the BESS can provide frequency-lower services in the FL, SL, and DL sub-markets when the charging binary variable equals $1$.

\subsection{Joint-Market Bidding Formulation} \label{subsec:model_joint}
\begin{figure}[!t]
    \centering
    \includegraphics[width=\linewidth]{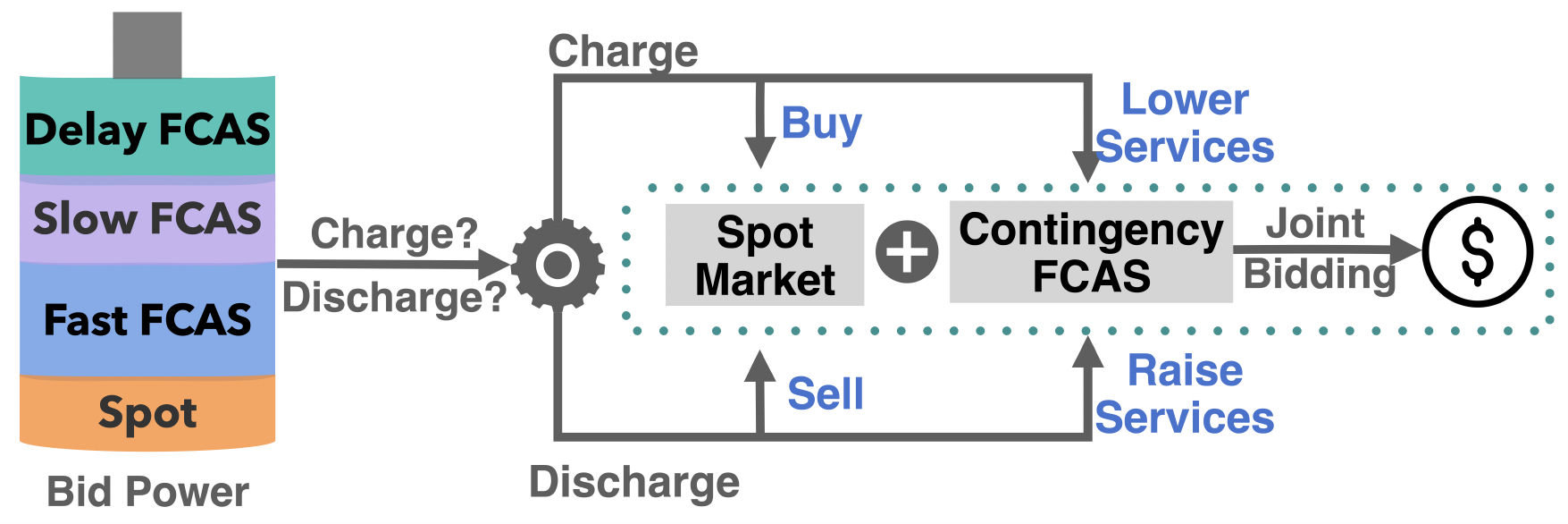}
    \caption{The system model paradigm.}
    \label{fig:system_model}
\end{figure}

Considering the multiple revenue streams of the BESS from the spot and contingency FCAS markets, we formulate the BESS joint bidding as an optimization problem, illustrated in Fig. \ref{fig:system_model}. The optimization objective is to maximize the total revenue deducted by BESS degradation cost as
\begin{equation}
    \label{eq:revenue_joint}
    \max \hspace{0.25em} R^\text{S} + R^\text{FCAS}-c\Delta t\sum_{t=1}^Tv_t^\text{dch}(p_t^\text{S}+p_t^\text{fast}+p_t^\text{slow}+p_t^\text{delay}),
\end{equation}
where the last term represents the degradation cost of the BESS. We model the degradation using discharge as it approximates the cycle~\mbox{\cite{anwar2022}}, and $c$ is a cost coefficient in AU\$/MWh.

Real-time dispatch of the BESS is constrained by its rated power formulated as
\begin{align}
    \label{eq:cons_power_FCAS_fast}
    0&\leq p_t^\text{fast}\leq P_\text{max}^\text{FCAS},\\
    \label{eq:cons_power_FCAS_slow}
    0&\leq p_t^\text{slow}\leq P_\text{max}^\text{FCAS},\\
    \label{eq:cons_power_FCAS_delay}
    0&\leq p_t^\text{delay}\leq P_\text{max}^\text{FCAS},\\
    \label{eq:cons_power_sum}
    0&\leq p_t^\text{S} + p_t^\text{fast} + p_t^\text{slow} + p_t^\text{delay} \leq P_\text{max},
\end{align}
where $P_\text{max}$ is the rated power (in MW) of the BESS and $P_\text{max}^\text{FCAS}$ is the maximum ancillary service power of the BESS (in MW) that is related to frequency deviation, frequency controller dead band, and the droop of the BESS~\cite{aemo2022_fcas}. The minimum allowable droop setting of any BESS with a nameplate rating of $5$ MW or above is $1.7\%$ in the NEM. 
Eq. \eqref{eq:cons_power_FCAS_fast} to \eqref{eq:cons_power_FCAS_delay} show that the bid power in the fast, slow, and delayed FCAS sub-markets must be within the maximum ancillary service power. Furthermore, Eq. \eqref{eq:cons_power_sum} represents that the sum of bid power in the spot and contingency FCAS markets cannot exceed the rated power of the BESS.

Also, charge and discharge operations of the BESS are also constrained by its current capacity $e_{t-1}+\Delta e_t$, where $e_{t-1}$ is its capacity after the previous dispatch interval and $\Delta e_t$ is the energy change in the current dispatch interval. The BESS's capacity must be within its lower and upper energy limits denoted by $E_\text{min}$ and $E_\text{max}$, which can be formulated as
\begin{equation}
    \label{eq:cons_energy}
    E_\text{min} \leq e_{t-1} + \Delta e_t \leq E_\text{max}.
\end{equation}
Both power exchange in the spot market and frequency service delivery in the contingency FCAS market lead to changes in stored energy of the BESS. The energy change caused by spot market particiation can be formulated as
\begin{equation}
    \label{eq:energy_change_spot}
    \Delta e_t^\text{S} = \Delta t \left(v_t^\text{ch}-v_t^\text{dch}\right)p_t^\text{S}.
\end{equation}

For contingency FCAS services, we introduce two contingency indicators $\mathbb{I}_t^\text{raise}$ and $\mathbb{I}_t^\text{lower}$ to represent the demand. Specifically, the raise/lower indicators are set to $1$ only on the occurrence of a major drop/raise of system frequency caused by a contingency event. The energy change from delivering frequency services can be formulated as
\begin{equation}
    \label{eq:energy_change_contingencyFCAS}
    \begin{aligned}
    \Delta e_t^\text{FCAS} &= \left(v_t^\text{ch}-v_t^\text{dch}\right)\left(\mathbb{I}_t^\text{raise}+\mathbb{I}_t^\text{lower}\right)\\
    &\hspace{1em}\times \left(\Delta t^\text{fast}p_t^\text{fast} +\Delta t^\text{slow}p_t^\text{slow}+\Delta t^\text{delay}p_t^\text{delay}\right),
    \end{aligned}
\end{equation}
where $\Delta t^\text{fast},\Delta t^\text{slow}, \Delta t^\text{delay}$ are the actual dispatch duration of the fast, slow, and delayed sub-markets. 
The summation of indicators in Eq. \eqref{eq:energy_change_contingencyFCAS}, i.e., $\mathbb{I}_t^\text{raise}+\mathbb{I}_t^\text{lower}$, represents the occurrence of a contingency event, while the last term represents the total energy output of three kinds of contingency FCAS services in one $5$-minute NEM dispatch interval. Combining energy changes from the spot and contingency FCAS markets, the total energy change $\Delta e_t$ can be expressed as 
$\Delta e_t = \Delta e_t^\text{S} + \Delta e_t^\text{FCAS}.$

\section{Methodology}\label{sec:method}
To solve the joint bidding problem, we first develop the transformer-based temporal feature extractor, namely the TTFE, in Section \ref{subsec:method_TTFE} to extract temporal information of multi-time-series energy prices in the spot and contingency FCAS markets, followed by Section \ref{subsec:method_DRL}, where we model the real-time bidding problem as an MDP and then develop the SAC~\cite{haarnoja2018} algorithm to learn an optimal joint-bidding strategy to maximize the overall revenue.

\subsection{Transformer-based Temporal Feature Extractor} \label{subsec:method_TTFE} 
We define a price vector consisting of market clearing prices in the spot and contingency FCAS markets as
\begin{equation}
\label{eq:TTFE_price_vec}
\bm{\rho}_t = \left[\rho_t^{\text{S}},\rho_t^\text{FR},\rho_t^\text{FL},\rho_t^\text{SR},\rho_t^\text{SL},\rho_t^\text{DR},\rho_t^\text{DL}\right].
\end{equation}
A temporal segment of length $L$ is further developed to store a series of historical price vectors, which can be defined as
\begin{equation}
\label{eq:TTFE_temporal_seg}
S_t = \left[\bm{\rho}_{t-L+1},\bm{\rho}_{t-L+2},\cdots,\bm{\rho}_{t}\right] \in \mathbb{R}^{L\times F},
\end{equation}
where $F$ is the feature dimension of the temporal segment, representing the number of participated markets.

To take advantage of historically multi-time-series energy prices, inspired by the transformer~\cite{vaswani2017} for its powerful capability in sequential feature extraction, we develop the TTFE to capture the temporal patterns of the temporal segment. Specifically, the proposed TTFE explores mutual influences of each price vector pair in the temporal segment context and assigns greater attention to price vectors with higher influence values (namely attention weights~\cite{vaswani2017}). Such "attention" information will be passed to the SAC algorithm (to be introduced in Section \ref{subsubsec:method_DRL_SAC}), thus enabling the BESS to be temporal-aware of volatile energy prices for better bidding decisions.

\begin{figure}[!t]
    \centering
    \includegraphics[width=\linewidth]{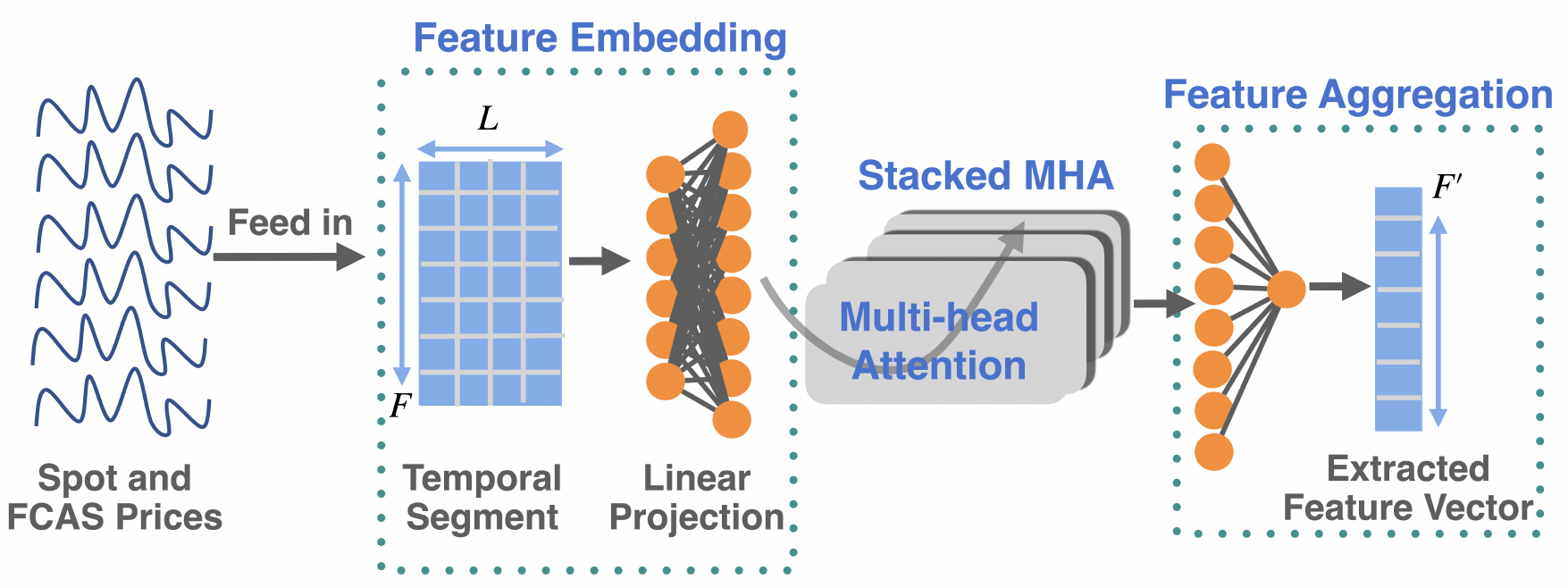}
    \caption{The framework of TTFE.}
    \label{fig:TTFE}
\end{figure}

The developed TTFE framework is illustrated in Fig. \ref{fig:TTFE}, including three key components: feature embedding, stacked multi-head attention (MHA), and feature aggregation. For each bidding decision making, we construct a temporal segment defined in Eq. \eqref{eq:TTFE_temporal_seg}, feed it into the TTFE, and derive an extracted feature vector to assist the SAC algorithm to bid in the spot and contingency FCAS markets.

\subsubsection{Feature Embedding} \label{subsubsec:method_TTFE_feature_embed}
Before fed into the stacked MHA, the input temporal segment first passes through the \textit{feature embedding} element for linear transformation (LT) in the feature space, which can be formulated as
\begin{equation}
\label{eq:TTFE_temporal_seg_trans}
S' = \text{LT}(S) = SW^\text{embed}+b^\text{embed}\in \mathbb{R}^{L\times F'},
\end{equation}
where $W^\text{embed}$ and $b^\text{embed}$ are the weighted and bias matrices of the LT layer, respectively. The transformed feature space dimension is denoted by $F'$.

\begin{figure}[!t]
    \centering
    \includegraphics[width=\linewidth]{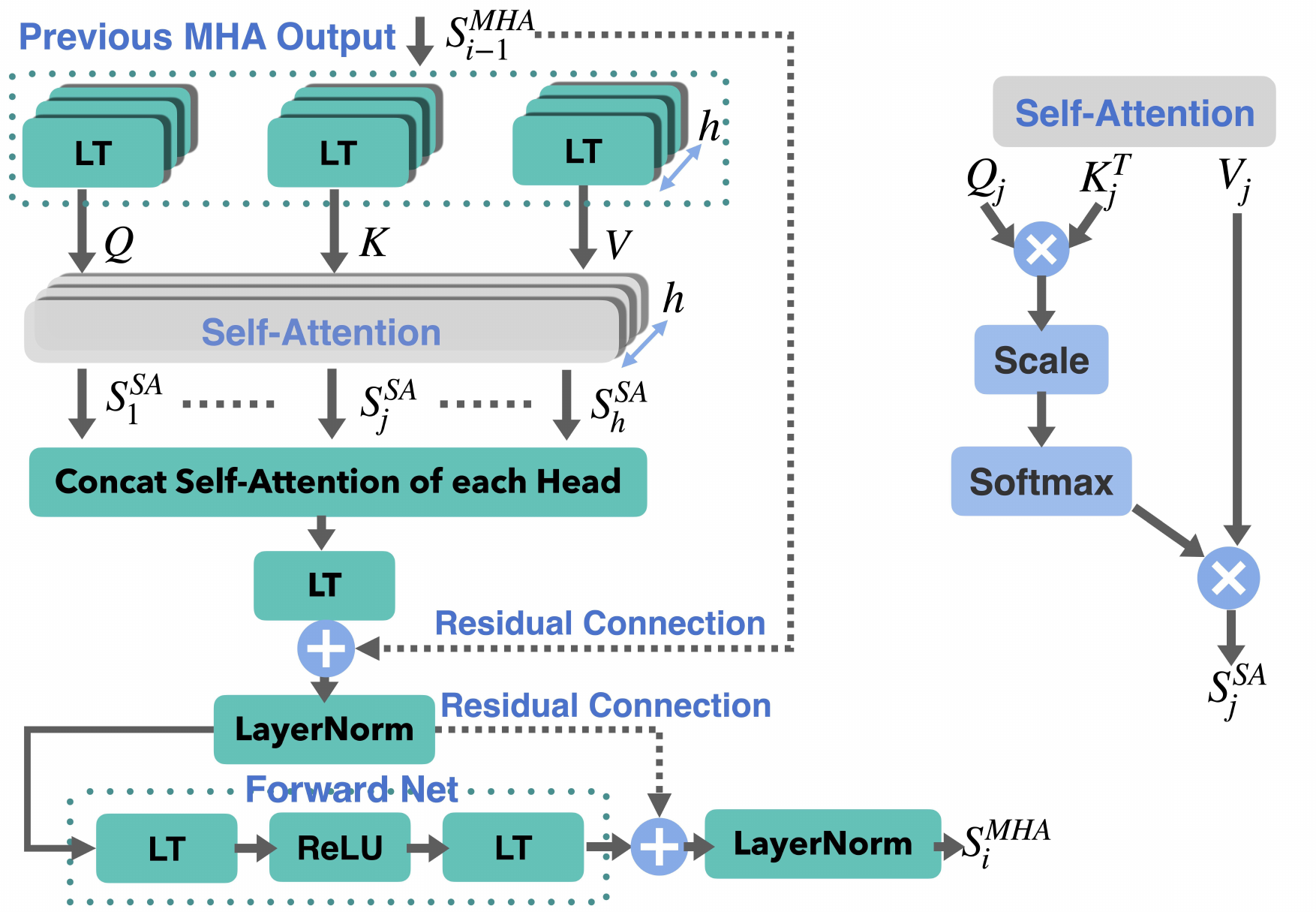}
    \caption{The internal structure of one MHA.}
    \label{fig:MHA}
\end{figure}

\subsubsection{Stacked MHA} \label{subsubsec:method_TTFE_stackedMHA}
The stacked MHA is the most critical component of our TTFE and plays a significant role in analyzing the temporal importance of historical prices. The internal structure of one MHA is illustrated in Fig. \ref{fig:MHA}. We denote the input of the $i$th MHA by $S_{i-1}^\text{MHA}$, where $i=1,\cdots,N_\text{MHA}$ is the index of the MHA and $N_\text{MHA}$ is the number of MHAs in the stacked MHA. In particular, the input of the first MHA $S_0^\text{MHA}$ is the transformed temporal segment $S'$. 

Here, we present the detailed temporal feature extraction process of the first MHA. To better explore temporal features of the transformed segment $S'$, the MHA splits the transformed feature space into multiple sub-spaces (namely the \textit{heads} of the MHA), aiming to learn various lower-scale feature maps as opposed to one all-encompassing map. The feature sub-space dimension of each MHA head is defined as $\frac{F'}{h}$, where $h$ is the number of MHA heads.

The transformed segment $S'$ passes through each MHA head three times in a parallel manner as shown in Fig. \ref{fig:MHA} to create the \textit{query}, \textit{key}, and \textit{value} matrices denoted by $Q$, $K$, and $V$, respectively. We formulate the simultaneous linear projection process as
\begin{align}
    Q_j = \text{LT}_j\left(S'\right) = S'W^Q_j \in \mathbb{R}^{L\times \frac{F'}{h}},\\
    K_j = \text{LT}_j\left(S'\right) = S'W^K_j \in \mathbb{R}^{L\times \frac{F'}{h}},\\
    V_j = \text{LT}_j\left(S'\right) = S'W^V_j \in \mathbb{R}^{L\times \frac{F'}{h}},
\end{align}
where $j$ is the index of the MHA head and $W^Q_j,W^K_j,W^V_j$ are weighted matrices.

The essence of the MHA is its self-attention (SA) mechanism whose structure is illustrated in detail on the right side of Fig. \ref{fig:MHA}. The SA calculates the aforementioned attention weights of each price vector pair, indicating the temporal significance of each price vector. We define one SA head as
\begin{equation}
    \label{eq:self_att}
    \begin{aligned}
    S_j^\text{SA}&=\text{SA}_j(Q,K,V)\\
    &=W_j^\text{att}V_j=\text{softmax}\left(\frac{Q_jK_j^T}{\sqrt{F'}}\right)V_j\in \mathbb{R}^{L\times \frac{F'}{h}},
    \end{aligned}
\end{equation}
where the $\text{softmax}(\cdot)$ function  is used for normalization. The result of the softmax operation is referred to as the attention matrix of the $j$th SA head denoted by $W^\text{att}_j\in\mathbb{R}^{L\times L}$, whose element $w^\text{att}_{m,n}$, i.e., the attention weight, represents the relative significance of how the price vector $\bm{\rho}_{t-L+n}$ affects $\bm{\rho}_{t-L+m}$ in the original temporal segment $S_t$. With the attention matrix, the MHA can pay greater attention to price vectors with larger attention weights. Moreover, the multiplication operation of $W^\text{att}_jV_j$ in Eq. \eqref{eq:self_att} aims to aggregate such ``attention'' information into the original temporal segment $S_t$, since the value matrix $V_j$ is obtained after several linear projections of $S_t$ and inherently saves its partial information.

To fuse the extracted temporal feature information in each MHA head, we then concatenate the outputs of each SA head and process it through one more LT layer as shown in Fig. \ref{fig:MHA}, which can be formulated as
\begin{equation}
    \hspace{-0.5em}\text{MHA}_i\left(Q,K,V\right)= \text{LT}\left(\text{Concat}\left(S_1^\text{SA},\cdots,S_h^\text{SA}\right)\right).
\end{equation}
The above LT output accumulates the initial input of the MHA (i.e., $S_{i-1}^\text{MHA}$) and passes through a \textit{LayerNorm} layer for normalization. Subsequently, we use the same \textit{Forward Net} structure in the original transformer model~\cite{vaswani2017}, i.e., two LT layers with the rectified linear unit (ReLU) as the activation function, to derive the final output of the MHA, which can be formulated as
\begin{equation}
\label{eq:attn_mat}
S_i^\text{MHA} = \text{ForwardNet}\left(\text{MHA}_i\left(Q,K,V\right)\right) \in \mathbb{R}^{L\times F'}.
\end{equation}

\subsubsection{Feature Aggregation} \label{subsubsec:model_TTFE_feature_agg}
It is challenging to integrate the stacked MHA output (i.e., $S_{N_\text{MHA}}^\text{MHA}$) into the DRL algorithm, due to its two-dimensional attribute and the considerably large amount of parameters involved. To address this, we introduce the one-dimensional global average pooling~\cite{lin2013} to compress the output of the stacked MHA, referred to as the \textit{Feature Aggregation} element in our TTFE. Such a pooling technique performs downsampling along the temporal dimension $L$. The final extracted feature vector of our proposed TTFE is formulated as
\begin{equation}
    \label{eq:extract_feature_vec}
    \bm{f} = \left[f_1,\cdots,f_n,\cdots,f_{F'}\right] \in \mathbb{R}^{1\times F'},
\end{equation}
with its element $f_n$ calculated via the global average pooling as $f_n = \frac{1}{L}\sum_{m=1}^{L} s_{m,n}$, where $s_{m,n}$ is the element of the stacked MHA's output $S_{N_\text{MHA}}^\text{MHA}$. The extracted feature vector carries temporal information of multi-time-series energy prices, sequentially fed to the following SAC algorithm for making better bidding decisions.

\subsection{Learning Optimal Joint-Bidding Strategy via DRL} \label{subsec:method_DRL}
\subsubsection{MDP Modeling} \label{subsubsec:method_DRL_MDP}
We model the consecutive BESS bidding problem as an MDP, consisting of four parts: state space $\mathbb{S}$, action space $\mathbb{A}$, probability space $\mathbb{P}$, and reward space $\mathbb{R}$.

\textbf{State Space} $\mathbb{S}$: The BESS's state aggregates the available price vector defined in Eq. \eqref{eq:TTFE_price_vec} and the extracted temporal feature vector derived in Eq. \eqref{eq:extract_feature_vec}, along with the BESS state of charge (SoC), which can be expressed as
\begin{equation}
    \label{eq:MDP_state}
    \bm{s}_t = \left[\text{SoC}_{t-1},\bm{\rho}_{t-1},\bm{f}_{t-1}\right],
\end{equation}
in which the SoC is defined as $\text{SoC}_{t-1} = \frac{e_{t-1}}{E}$, where $E$ is the storage capacity of the BESS. 

\textbf{Action Space} $\mathbb{A}$: Actions of the BESS include charge/discharge variables $v_t^\text{ch},v_t^\text{dch}$ and bids in the spot and contingency FCAS markets, which are scaled by $P_\text{max}$. Thus, action in the spot market $a_t^\text{S}$ falls into the range from $0$ to $1$, while actions in the fast, slow, and delayed FCAS sub-markets $a_t^\text{fast}$, $a_t^\text{slow}$, and $a_t^\text{delay}$ are from $0$ to $\frac{P_\text{max}^\text{FCAS}}{P_\text{max}}$. Actions of the BESS are defined as
\begin{equation}
    \label{eq:MDP_action}
    \bm{a}_t = \left[v_t^\text{dch},v_t^\text{ch},a_t^\text{S},a_t^\text{fast},a_t^\text{slow},a_t^\text{delay}\right].
\end{equation}

\textbf{Probability Space} $\mathbb{P}$: The probability space refers to the probability set of transitioning to the next state after taking a deterministic action, which is defined as $\mathbb{P}\left(\bm{s}_{t+1}|\bm{s}_t,\bm{a}_t\right)$.

\textbf{Reward Space} $\mathbb{R}$: The BESS receives a reward after taking action $\bm{a}_t$ at state $\bm{s}_t$, which reflects the effectiveness of the bidding decision. Thus, designing an appropriate reward function is of great importance to facilitate the BESS to make better bidding decisions.

For the spot market, we introduce two charge/discharge indicators denoted by $\mathbb{I}^\text{ch}_t$ and $\mathbb{I}^\text{dch}_t$, formulated as
\begin{equation}
    \label{eq:dch_ch_indicators}
    \mathbb{I}_t^\text{ch} = \text{sgn}\left(\bar{\rho}_t^\text{S}-\rho_t^\text{S}\right),\quad \mathbb{I}_t^\text{dch} = \text{sgn}\left(\rho_t^\text{S}-\bar{\rho}_t^\text{S}\right),
\end{equation}
where $\text{sgn}(\cdot)$ is the sign function and $\bar{\rho_t}$ is the exponential moving average of the spot price~\cite{wang2018}, which is defined as
\begin{equation}
    \label{eq:moving_avg_spot_price}
    \bar{\rho}_t^\text{S} = \tau^\text{S}\bar{\rho}_{t-1}^\text{S} + (1-\tau^\text{S})\rho_t^\text{S},
\end{equation}
where $\tau^\text{S}\in(0,1)$ is a smoothing parameter. The charge and discharge indicators incentivize the BESS to buy low ($\rho_t^\text{S}<\bar{\rho}_t^\text{S}$) and sell high ($\rho_t^\text{S}>\bar{\rho}_t^\text{S}$). Any bids violating such an arbitrage guideline will be penalized. Hence, the spot market rewards are formulated as
\begin{equation}
    \label{eq:reward_spot}
    \begin{aligned}
    r_t^\text{S} &= a_t^\text{S}\rho_t^\text{S}\left(v_t^\text{dch}\eta^\text{dch}-v_t^\text{ch}\frac{1}{\eta^\text{ch}}\right)\\
    &+\beta^\text{S}a_t^\text{S}|\rho_t^\text{S}-\bar{\rho}_t^\text{S}|\left(\mathbb{I}_t^\text{dch}v_t^\text{dch}\eta^\text{dch}+\mathbb{I}_t^\text{ch}v_t^\text{ch}\frac{1}{\eta^\text{ch}}\right),
    \end{aligned}
\end{equation}
where the first term represents the inherent revenue gain/loss caused by the power exchange in the spot market. The second term is the reward for performing energy arbitrage based on the \textit{buy-low-sell-high} guideline, with the coefficient $\beta^\text{S}$ indicating its importance.

Moreover, rewards from the fast (FR and FL), slow (SR and SL), and delayed (DR and DL) contingency FCAS sub-markets can be expressed as

\begin{align}
    \label{eq:reward_fast}
    r_t^\text{fast} &= a_t^\text{fast}\left(v_t^\text{dch}\eta^\text{dch}\rho_t^\text{FR}+v_t^\text{ch}\frac{1}{\eta^\text{ch}}\rho_t^\text{FL}\right),\\
    \label{eq:reward_slow}
    r_t^\text{slow} &= a_t^\text{slow}\left(v_t^\text{dch}\eta^\text{dch}\rho_t^\text{SR}+v_t^\text{ch}\frac{1}{\eta^\text{ch}}\rho_t^\text{SL}\right),\\
    \label{eq:reward_delay}
    r_t^\text{delay} &= a_t^\text{delay}\left(v_t^\text{dch}\eta^\text{dch}\rho_t^\text{DR}+v_t^\text{ch}\frac{1}{\eta^\text{ch}}\rho_t^\text{DL}\right).
\end{align}

Combining all the rewards from the spot and contingency FCAS markets, we obtain the reward function of the BESS as
\begin{equation}
    \label{eq:reward}
    r_t = r_t^\text{S} + r_t^\text{fast} + r_t^\text{slow} + r_t^\text{delay}.
\end{equation}

Additionally, a constant penalty, e.g., $50$ used in our algorithm, is added to the reward $r_t$ when the BESS's bid violates its energy limits. Such a penalty often comes along with the termination of the training episode, informing the RL agent not to violate environment constraints for the aim of reward maximization. Given that the constant penalty cannot perfectly avoid constraint violation, we have also clipped the BESS's bidding decisions within its battery SoC limits, preventing the BESS from violating both its physical constraints and market rules.

\subsubsection{Optimizing MDP by SAC} \label{subsubsec:method_DRL_SAC}
We employ a state-of-the-art DRL algorithm, namely SAC~\cite{haarnoja2018}, to optimize the derived MDP. SAC aims to learn an optimal bidding strategy denoted by $\pi(\bm{a}_t|\bm{s}_t)$ that maximizes the expected returns over the total time frame, which can be formulated as
\begin{equation}
    \label{eq:SAC_obj}
    J_\pi = \mathbb{E}_{\bm{s}_t\sim\mathbb{P},\bm{a}_t\sim\pi(\bm{s}_t)} \left[\sum_{t=1}^T \gamma^{t-1} r_t \right],
\end{equation}
where $\gamma$ is the discounted factor.

SAC follows an actor-critic framework, where the actor, i.e., the bidding strategy $\pi$, determines an action based on the current state, and the critic consists of two functions: the value function $V(\bm{s}_t)$ and the Q function $Q(\bm{s}_t,\bm{a}_t)$, which can examine the effectiveness of the state-action pair. The Q function can be formulated using the Bellman equation as
\begin{equation}
    \label{eq:Bellman_eq}
    Q(\bm{s}_t,\bm{a}_t) = r_t + \gamma \mathbb{E}_{\bm{s}_{t+1}\sim \mathbb{P}}\left[V\left(\bm{s}_{t+1}\right)\right],
\end{equation}
with the value function defined as
\begin{equation}
    \label{eq:value_func}
    V(\bm{s}_t) = \mathbb{E}_{\bm{a}_t\sim\pi(\bm{s}_t)}\left[Q(\bm{s}_t,\bm{a}_t)\right].
\end{equation}

The essence of the SAC is to introduce an entropy term for the bidding strategy, which can improve its stochasticity and better explore the underlying bidding decision space. The entropy term is defined as
\begin{equation}
    \label{eq:entropy}
    \mathcal{H}\left[\pi\right] = -\mathbb{E}_{\bm{s}_t\sim\mathbb{P}, \bm{a}_t\sim\pi(\bm{s}_t)}\left[\log\pi\left(\bm{a}_t|\bm{s}_t\right)\right].
\end{equation}
Entropy is originally defined as a measure of information content given a distribution. In the context of the adopted SAC algorithm, the entropy term describes the diversity of the output actions, ensuring diversification of the action policy, i.e., increasing the information content of the action distribution, thereby preventing the policy from becoming stuck in local minima. Though the entropy term introduces aleatoric uncertainty in the DRL-based bidding strategy~\cite{lockwood2022_uncertaintyRL}, it enables the action policy to strike a more effective balance between reward and entropy maximization, thereby approximating the global optimum.

With the addition of entropy, the objective of the bidding strategy and value function can be rewritten as
\begin{align}
    \label{eq:SAC_obj_rewrite}
    J_\pi &= \mathbb{E}_{\bm{s}_t\sim \mathbb{P},\bm{a}_t\sim \pi(\bm{s}_t)} \left[\sum_{t=1}^T \gamma^{t-1} r_t+\alpha\mathcal{H}\left[\pi\right]\right],\\
    \label{eq:value_func_rewrite}
    V(\bm{s}_t) &= \mathbb{E}_{\bm{a}_t\sim\pi(\bm{s}_t)}\left[Q(\bm{s}_t,\bm{a}_t)-\alpha\log\pi(\bm{a}_t|\bm{s}_t)\right],
\end{align}
where $\alpha$ is the temperature parameter representing the relative significance of the entropy against the returns.

We can find a better bidding strategy in SAC based on
\begin{equation}
    \label{eq:action_straetgy_update_policy}
    \pi_\text{new} = \arg\min_{\pi'} \text{D}_\text{KL}\left[
    \pi'(\bm{s}_t)
    \left|\right|
    \frac{\exp\left[Q\left(\bm{s}_t,\pi\left(\bm{s}_t\right)\right)\right]}{Z(\bm{s}_t)}
    \right],
\end{equation}
where the Kullback-Leibler (KL) divergence $\text{D}_\text{KL}$ measures the distribution distance between the update strategy candidate $\pi'$ and the exponential Q function normalized by its partition function $Z(\bm{s}_t)$. The partition function normalizes the Q function by enumerating all input states in the state space $\mathbb{S}$, which does not affect the gradient to the new policy and can be ignored during the training process of the action policy.

We apply neural networks as function approximators to estimate the bidding strategy $\pi_\phi(\bm{a}_t|\bm{s}_t)$, the value function $V_\psi(\bm{s}_t)$, and the Q function $Q_\theta(\bm{s}_t,\bm{a}_t)$, where $\phi$, $\psi$, $\theta$ are their corresponding neural network parameters. The Adam gradient descent optimizer~\cite{haarnoja2018} is used to train these neural networks. 

\textbf{Update $\pi_\phi$}: The bidding strategy is updated by minimizing the aforementioned KL divergence, whose gradient can be formulated as 
\begin{equation}
    \label{eq:gradient_pi}
    \hspace{-0.5em}\nabla_\phi J_\pi(\phi) = \nabla_\phi \mathbb{E}_{\bm{s}_t\sim \mathbb{B}} \left[
    \text{D}_\text{KL}\left[
    \pi_\phi
    \left|\right|
    \frac{\exp\left[Q_\theta\left(\bm{s}_t,\pi_\phi\right)\right]}{Z_\theta(\bm{s}_t)}
    \right]
    \right],
\end{equation}
where the replay buffer $\mathbb{B}$ is introduced to store transitions defined as $\{\bm{s}_t,\bm{a}_t,r_t,\bm{s}_{t+1}\}$ in the MDP. To ensure that the sum of bid power does not exceed the rated power of the BESS as defined in Eq. \eqref{eq:cons_power_sum}, we develop an ancillary loss function for the bidding strategy objective, which can be formulated as
\begin{equation}
    \label{eq:loss_func_pi}
    L_\pi(\phi) = a_t^\text{bid} \mathbb{I}\left(a_t^\text{bid} > 1\right),
\end{equation}
with the sum of bids $a_t^\text{bid}$ defined as
\begin{equation}
    \label{eq:bid_sum}
    a_t^\text{bid} = a_t^\text{S} + a_t^\text{fast} + a_t^\text{slow} + a_t^\text{delay}.
\end{equation}
Hence, the gradient of the bidding strategy in Eq. \eqref{eq:gradient_pi} can be rewritten as
\begin{equation}
    \label{eq:gradient_pi_final}
    \nabla_\phi J_\pi(\phi) \leftarrow \nabla_\phi J_\pi(\phi) + \beta^L \nabla_\phi \mathbb{E}_{\bm{a}_t\sim\mathbb{B}}\left[L_\pi(\phi)\right],
\end{equation}
where $\beta^L$ is the coefficient of the proposed loss function.

\textbf{Update $V_\psi$}: We update the value network by minimizing the squared residual error of Eq. \eqref{eq:value_func_rewrite}. The gradient of the value network objective is formulated as
\begin{equation}
    \label{eq:gradient_V}
    \nabla_\psi J_V(\psi) = \nabla_\psi \mathbb{E}_{\bm{s}_t\in \mathbb{B}}
    \left[\frac{1}{2}\left[V_\psi(\bm{s}_t)-\hat{V}(\bm{s}_t)\right]^2\right],
\end{equation}
with the estimated value function $\hat{V}(\bm{s}_t)$ expressed as
\begin{equation}
    \hat{V}(\bm{s}_t) = \mathbb{E}_{\bm{a}_t\sim\pi_\phi(\bm{s}_t)}\left[Q_\theta(\bm{s}_t,\bm{a}_t) -\alpha\log\pi_\phi(\bm{a}_t|\bm{s}_t) \right].
\end{equation}

\textbf{Update $Q_\theta$}: Similarly, the Q network is optimized by minimizing the residual error of the Bellman equation defined in Eq. \eqref{eq:Bellman_eq}. To stabilize the optimization process, a target value network $V_{\hat{\psi}}$ is adopted to estimate the Q function, i.e., the right-hand side of Eq. \eqref{eq:Bellman_eq}. The gradient of the Q network objective is formulated as
\begin{equation}
    \label{eq:approx_Q}
    \hspace{-0.4em}\nabla_\theta J_Q(\theta) = \nabla_\theta \mathbb{E}_{\bm{s}_t\sim \mathbb{B}}
    \left[
    \frac{1}{2}
    \left[
    Q_\theta\left(\bm{s}_t,\pi_\phi\right)-
    \hat{Q}(\bm{s}_t,\pi_\phi)
    \right]^2
    \right],
\end{equation}
with the estimated Q function $\hat{Q}(\bm{s}_t,\pi_\phi)$ expressed as
\begin{equation}
\label{eq:target_value_net}
    \hat{Q}(\bm{s}_t,\pi_\phi) = r_t + \gamma\mathbb{E}_{\bm{s}_{t+1}\in\mathbb{P}}
    \left[
    V_{\hat{\psi}}(\bm{s}_{t+1})
    \right].
\end{equation}
Making use of the value network parameters $\psi$, we update the target value network in an exponential moving average manner, formulated as
\begin{equation}
    \label{eq:target_net_update}
    \hat{\psi} \leftarrow \tau^\psi\psi + (1-\tau^\psi)\hat{\psi},
\end{equation}
where $\tau^\psi\in(0,1)$ is the smoothing parameter. 

The detailed algorithmic procedure of our TempDRL is presented in Algorithm \ref{algo}.
\begin{algorithm}[!t]
\caption{The TempDRL Bidding Strategy}
\label{algo}
\begin{algorithmic}
\STATE Initialize parameters of the TTFE and SAC.
\FOR{$t=1,\cdots,T$}
\STATE Construct and feed the temporal segment $S_{t-1}$ into the TTFE; Derive the extracted feature vector $\bm{f}_{t-1}$.
\STATE Get the current state $\bm{s}_t=\left[\text{SoC}_{t-1},\bm{\rho}_{t-1},\bm{f}_{t-1}\right]$
\STATE Get action $\bm{a}_t=\pi_\phi(\bm{s}_t)$ and reward $r_t$.
\IF{action violates the BESS energy limits}
\STATE $\bm{a}_t \leftarrow \bm{0}$.
\ENDIF
\STATE Transit into the next state $\bm{s}_{t+1}$ via $\mathbb{P}\left(\bm{s}_{t+1}|\bm{s}_t,\bm{a}_t\right)$.
\STATE Store transition $\{\bm{s}_t,\bm{a}_t,r_t,\bm{s}_{t+1}\}$ into replay buffer $\mathbb{B}$.
\STATE Update parameters with sufficient transitions.
\ENDFOR
\end{algorithmic}
\end{algorithm}

\begin{table}[!t]
    \centering
    \caption{The initialized parameters.}
    \begin{tabular}{cc||cc||cc}
    \hline
    $\eta^\text{ch},\eta^\text{dch}$ & $0.95$ & $\Delta t$ & $5$ mins & $P_\text{max}$ & $2$ MW\\
    $P_\text{max}^\text{FCAS}$ & $1$ MW & $E$ & $10$ MWh & $E_\text{min}$ & $0.5$ MWh\\
    $E_\text{max}$ & $9.5$ MWh & $\Delta t^\text{fast}$ & $6$ secs & $\Delta t^\text{slow}$ & $55$ secs\\
    $\Delta t^\text{delay}$ & $4$ mins & $L$ & $32$ & $F$ & $7$\\
    $h$ & $8$ & $F'$ & $64$ & $N_\text{MHA}$ & $2$\\
    $\tau^\text{S}$ & $0.9$ & $\beta^\text{S}$ & $10$ & $\gamma$ & $0.99$\\
    $\beta^L$ & $10$ & $\tau^\psi$ & $0.01$ & $c$ & AU\$$1$/MWh \\ 
    $\eta^\pi$ & $0.0003$ & $\eta^V$ & $0.0003$ & $\eta^Q$ & $0.0003$\\
    \hline
    \end{tabular}
    \label{tab:parameters}
\end{table}

In summary, the gradient descent processes for the bidding strategy, value network, and Q network are formulated as
\begin{align}
    \label{eq:gradient_descent_pi}
    \phi \leftarrow \phi - \eta^\pi \nabla_\phi J_\pi(\phi),\\
    \label{eq:gradient_descent_V}
    \psi \leftarrow \psi - \eta^V \nabla_V J_V(\psi),\\
    \label{eq:gradient_descent_Q}
    \theta \leftarrow \theta - \eta^Q \nabla_Q J_Q(\theta),
\end{align}
where $\eta^\pi,\eta^V,\eta^Q$ are corresponding learning rates. As the TTFE is integrated into the SAC algorithm and shared by the bidding strategy $\pi_\phi$, the Q network $Q_\theta$, and the value network $V_\psi$, the TTFE is simultaneously updated during the gradient descent process.

\section{Experiments and Results}\label{sec:exp}
\subsection{Experimental Settings} \label{subsec:exp_setting}
Our proposed TempDRL method is trained and evaluated using energy prices from the five jurisdictions of the NEM in 2016, including Victoria (VIC), New South Wales (NSW), Queensland (QLD), South Australia (SA), and Tasmania (TAS). The energy prices from the first ten months are utilized for training while the last two months are reserved for evaluation. The length of one training episode is one day, consisting of $288$ time frames, given that the NEM dispatch interval is five minutes. The training of the DRL algorithms is carried out using an Nvidia TITAN RTX graphics processing unit. The initialized parameters of the TempDRL are provided in Table \ref{tab:parameters}. Moreover, the dimensionality of LT layers in the TTFE (except the first LT layer in the \textit{Forward Net}) is $64$, while the dimensionality of the first LT layer in the \textit{Forward Net} is set as $2048$. The MLP parts of the policy, Q, and value neural networks all have two hidden layers (i.e., two LT layers) with a dimensionality of $512$ for each layer. The activation function used in the SAC algorithm is the ReLU function. Notably, the outputs of the MLP are finally processed by one more hyperbolic tangent activation function to constrain the output range into $[-1,1]$. 

For the battery's parameters, the charging/discharging efficiencies, i.e., $\eta^\text{ch}$ and $\eta^\text{dch}$, are both $0.95$. The BESS's storage capacity is $10$ MWh, with the rated power and maximum ancillary service power set as $2$ MW and $1$ MW, respectively. The minimum and maximum energy limits of the BESS are set as $0.5$ MWh and $9.5$ MWh, respectively, equivalent to $5\%$ and $95\%$ SoC limits.

Three scenarios are analyzed in which the BESS participates: 1) the spot market only; 2) the contingency FCAS market only; and 3) both markets jointly.

\begin{table*}[!t]
    \centering
    \caption{The evaluation revenue of the TempDRL and benchmarks in five jurisdictions of the NEM (Unit: AU\$).}
    \begin{tabular}{c|c|c|c|c|c|c|c|c}
    \hline 

    \multirow{2}{*}{State} & \multirow{2}{*}{Market} & \multicolumn{4}{|c|}{Evaluation Revenue} & \multicolumn{3}{|c}{Revenue Boosts of TempDRL} \\ 

    \cline{3-9}

    && MLP-DRL & LP\&O & TP\&O & TempDRL & MLP-DRL & LP\&O & TP\&O \\ 

    \hline 

    \multirow{3}{*}{VIC} & Spot & $19,435$ & $24,547$ & $25,326$ & $\bm{30,467}$ & $57\%$ & $24\%$ & $20\%$ \\ 

    \cline{2-9}
    
    & FCAS & $7,840$ & $7,824$ & $8,296$ & $\bm{9,623}$ & $23\%$ & $23\%$ & $16\%$ \\ 
    
    \cline{2-9}
    
    & Joint & $22,975$ & $29,049$ & $30,320$ & $\bm{35,686}$ & $55\%$ & $23\%$ & $18\%$ \\

    \hline 

    \multirow{3}{*}{NSW} & Spot & $14,817$ & $17,297$ & $18,341$ & $\bm{22,853}$ & $54\%$ & $32\%$ & $25\%$ \\
    
    \cline{2-9}
    
    & FCAS & $8,065$ & $7,983$ & $7,338$ & $\bm{10,143}$ & $26\%$ & $27\%$ & $38\%$ \\
    
    \cline{2-9}
    
    & Joint & $18,052$ & $19,564$ & $19,682$ & $\bm{25,948}$ & $44\%$ & $33\%$ & $32\%$ \\
    
    \hline 
    
    \multirow{3}{*}{QLD} & Spot & $28,532$ & $33,873$ & $32,877$ & $\bm{40,625}$ & $42\%$ & $20\%$ & $24\%$ \\
    
    \cline{2-9}
    
    & FCAS & $6,420$ & $7,704$ & $7,848$ & $\bm{10,583}$ & $65\%$ & $37\%$ & $35\%$ \\
    
    \cline{2-9}
    
    & Joint & $31,052$ & $38,381$ & $39,640$ & $\bm{46,703}$ & $50\%$ & $22\%$ & $18\%$ \\
    
    \hline 
    
    \multirow{3}{*}{SA} & Spot & $36,334$ & $37,200$ & $32,892$ & $\bm{44,483}$ & $22\%$ & $20\%$ & $35\%$ \\
    
    \cline{2-9}
    
    & FCAS & $10,186$ & $10,953$ & $10,330$ & $\bm{13,456}$ & $32\%$ & $23\%$ & $30\%$ \\
    
    \cline{2-9}
    
    & Joint & $40,920$ & $41,413$ & $41,165$ & $\bm{49,146}$ & $20\%$ & $19\%$ & $19\%$ \\
    
    \hline 
    
    \multirow{3}{*}{TAS} & Spot & $30,812$ & $33,532$ & $32,240$ & $\bm{41,032}$ & $33\%$ & $22\%$ & $27\%$ \\
    
    \cline{2-9}
    
    & FCAS & $26,951$ & $22,618$ & $23,468$ & $\bm{31,877}$ & $18\%$ & $41\%$ & $36\%$ \\
    
    \cline{2-9}
    
    & Joint & $45,805$ & $43,287$ & $44,558$ & $\bm{52,478}$ & $15\%$ & $21\%$ & $18\%$ \\
    
    \end{tabular}
    \label{tab:benchmark_comp}
\end{table*}

\subsection{Benchmark Comparisons} \label{subsec:exp_TTFE}
\subsubsection{Evaluation Revenue Comparisons}
To assess the efficacy of the proposed TTFE, we train and evaluate our TempDRL model with and without TTFE. Additionally, we establish a \textit{predict-and-optimize} (P\&O) benchmark, i.e., a deterministic model predictive control (DMPC)-based approach~\cite{badoual2021_DMPC}, for comparison purposes. The P\&O method involves forecasting market clearing prices via a long short-term memory (LSTM) network and solving the joint bidding problem through mixed integer linear programming, which is implemented via the PuLP library~\cite{mitchell2011} and solved by the built-in Gurobi solver. The look-ahead period for the LSTM model is set as $48$, i.e., predicting energy prices in the next $48$ dispatch intervals. For fair comparisons, we also incorporate the transformer model into the P\&O framework. Specifically, the extracted feature vector (defined in \eqref{eq:extract_feature_vec}) is directly fed into an LT layer for forecasting. We refer to the P\&O benchmark with the \textbf{L}STM and the \textbf{t}ransformer model as LP\&O and TP\&O, respectively. Moreover, the DRL-based bidding strategy without the TTFE (i.e., only with MLPs) is referred to as MLP-DRL.

The results based on the revenue derived in VIC and the other four states (including NSW, QLD, SA, and TAS) are presented in Fig. \ref{fig:benchmark_comp_VIC} and Fig. \ref{fig:benchmark_comp_NSW_QLD_SA_TAS} (in Appendix A), respectively. The detailed results, including revenue boosts compared to the three benchmarks, for all five jurisdictions in the NEM are provided in Table \ref{tab:benchmark_comp}.

The results of our analysis, as depicted in Fig. \ref{fig:benchmark_comp_MLPDRL_VIC} and \ref{fig:benchmark_comp_TempDRL_VIC}, indicate that joint bidding consistently leads to higher revenue compared to participating in individual markets in VIC, so as other four jurisdictions, as shown in Table \ref{tab:benchmark_comp}. This revenue increase can be attributed to BESS's capability to fully utilize its potential and capitalize on the flexibility offered by both the spot and contingency FCAS markets.

\begin{figure}[!t]
    \centering
    \subfloat[VIC -- MLP-DRL]{
    \includegraphics[width=0.48\linewidth]{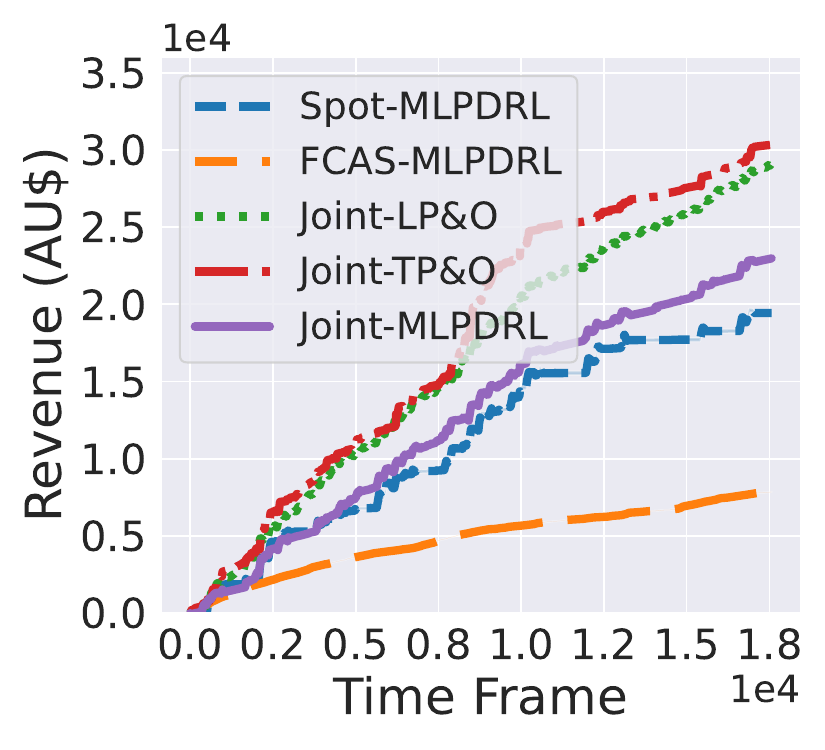}
    \label{fig:benchmark_comp_MLPDRL_VIC}
    }
    \subfloat[VIC -- TempDRL]{
    \includegraphics[width=0.48\linewidth]{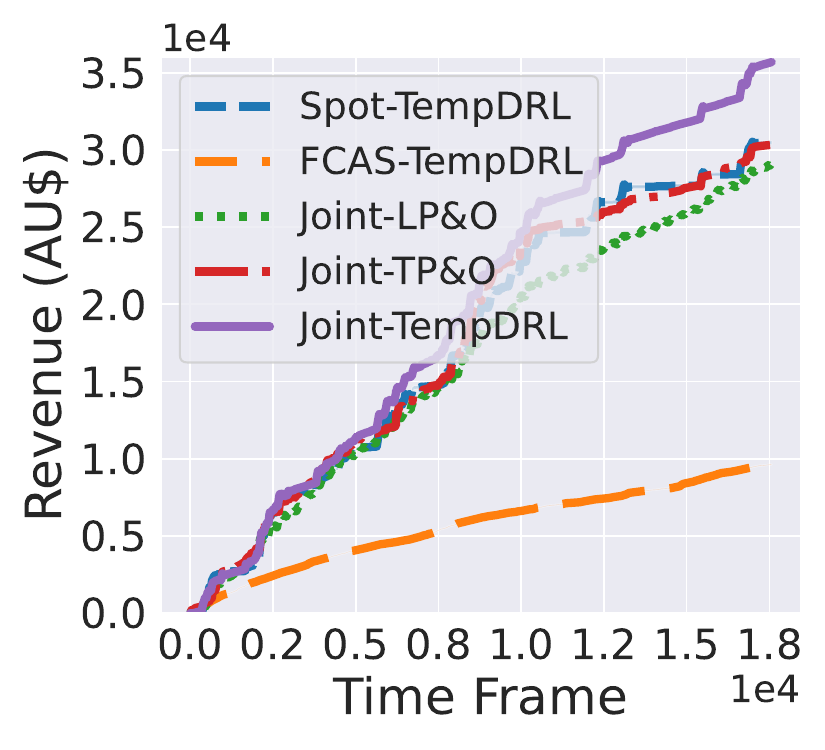}
    \label{fig:benchmark_comp_TempDRL_VIC}
    }
    \caption{Revenue comparisons of the TempDRL method with benchmarks in spot, contingency FCAS, and joint markets of VIC.}
    \label{fig:benchmark_comp_VIC}
\end{figure}

\begin{figure}[!t]
    \centering
    \subfloat[Spot.]{
    \includegraphics[width=.48\linewidth]{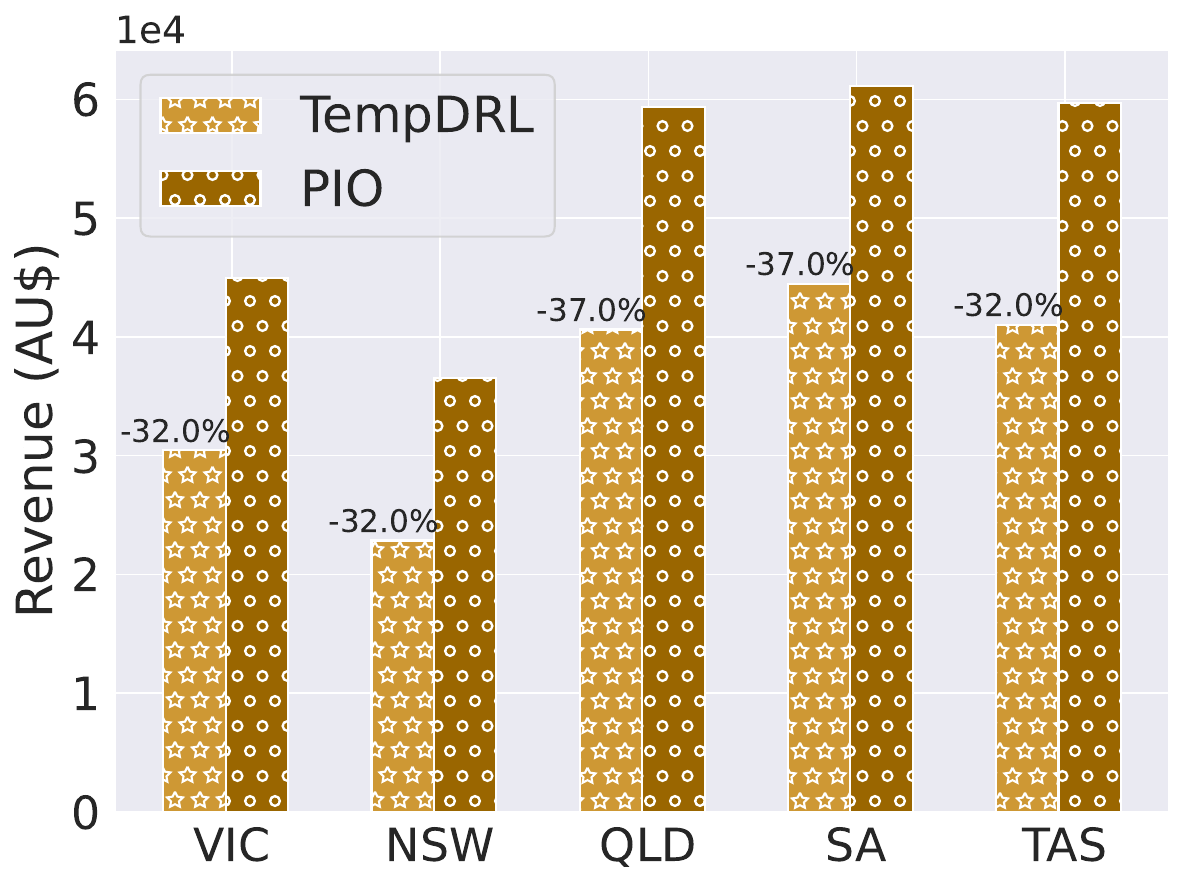}
    \label{fig:benchmark_comp_tempDRL_PIO_spot}
    }
    \subfloat[Contingency FCAS.]{
    \includegraphics[width=.48\linewidth]{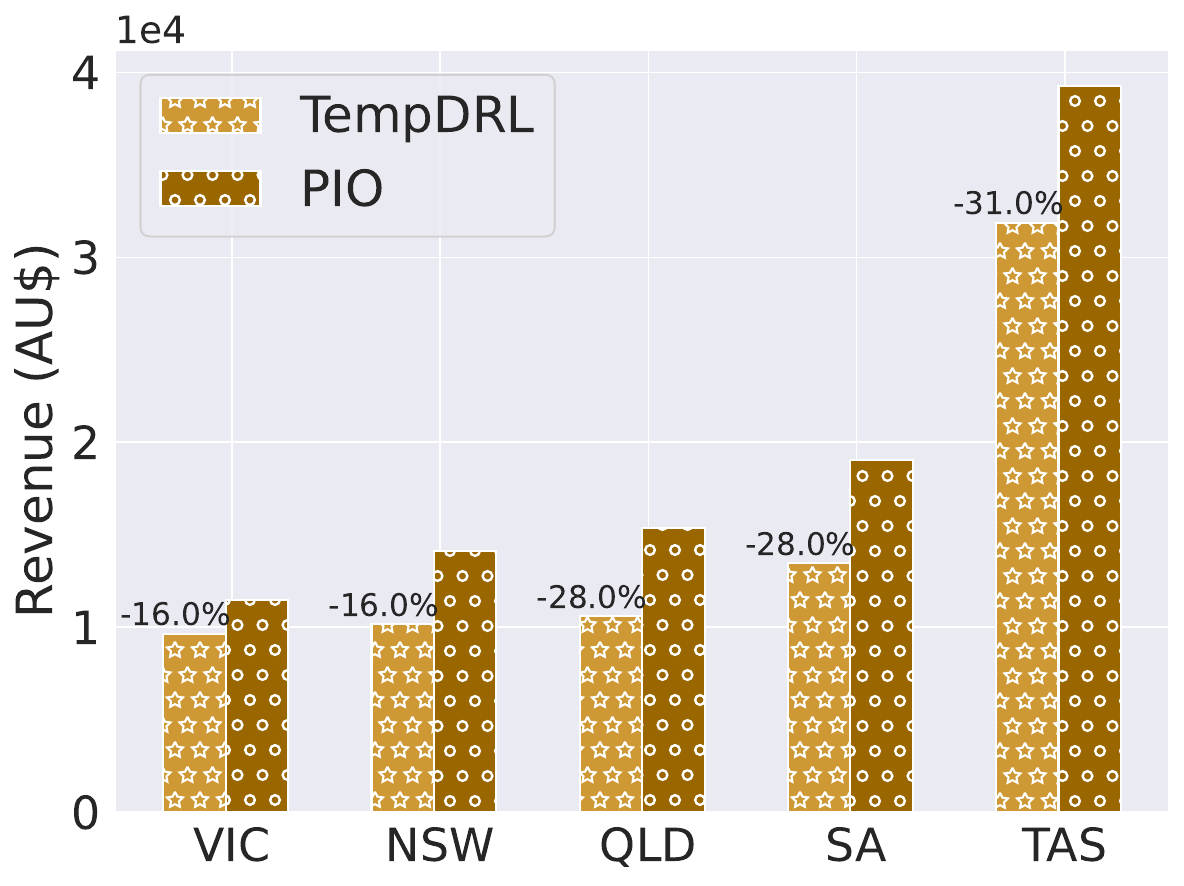}
    \label{fig:benchmark_comp_tempDRL_PIO_CFCAS}
    }
    \caption{Revenue comparisons of the TempDRL and the PIO in spot and contingency FCAS markets at five jurisdictions of the NEM.}
    \label{fig:benchmark_comp_tempDRL_PIO}
\end{figure}

\begin{figure}[!t]
    \centering
    \subfloat[Spot.]{
    \includegraphics[width=.48\linewidth]{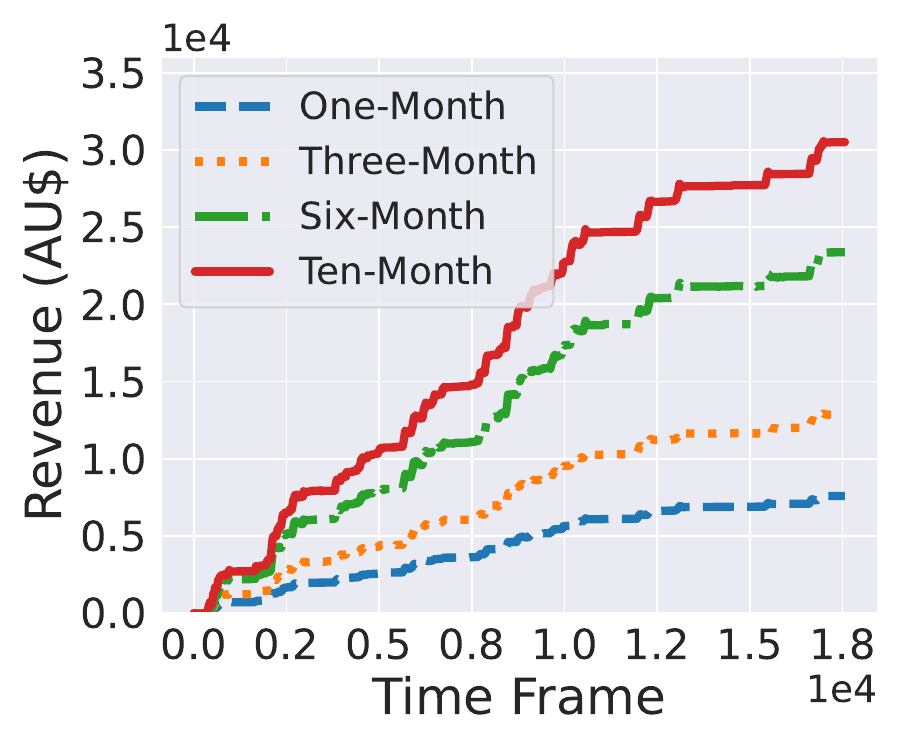}
    \label{fig:train_size_comp_TempDRL_VIC_ES}
    }
    \subfloat[Contingency FCAS.]{
    \includegraphics[width=.48\linewidth]{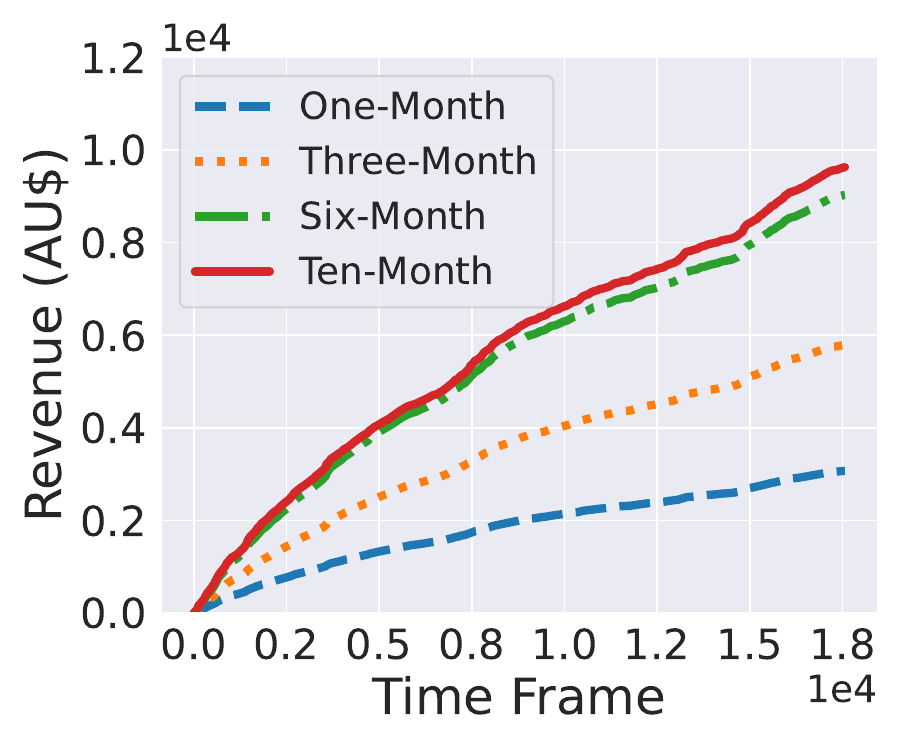}
    \label{fig:train_size_comp_TempDRL_VIC_Contingency_FCAS}
    }
    \caption{Revenue comparisons of the TempDRL trained with different sizes of dataset in the spot and contingency FCAS market of VIC.}
    \label{fig:train_size_comp_TempDRL_VIC}
\end{figure}

More importantly, the results demonstrate that introducing the TTFE can substantially improve bidding performance in all three bidding scenarios, as shown in Table \mbox{\ref{tab:benchmark_comp}}.  What stands out in both Table \mbox{\ref{tab:benchmark_comp}} and Fig. \mbox{\ref{fig:benchmark_comp_VIC}} is the significant revenue boost after introducing the TTFE in the joint market, i.e., our TempDRL approach shown in the purple solid line in Fig. \mbox{\ref{fig:benchmark_comp_TempDRL_VIC}}. This considerable improvement has surpassed both the LP\&O and the TP\&O benchmarks (shown in the green dotted and red dash-dot lines, respectively) by approximately $23\%$ and $18\%$ in VIC, equivalent to AU\$$6,638$ and AU\$$5,366$ in total, respectively.

More importantly, the results demonstrate that introducing the TTFE can substantially improve bidding performance in all three bidding scenarios, as shown in Table \ref{tab:benchmark_comp}. What stands out in both Table \ref{tab:benchmark_comp} and Fig. \ref{fig:benchmark_comp_VIC} is the significant revenue boost after introducing the TTFE in the joint market, i.e., our TempDRL approach shown in the purple solid line in Fig. \ref{fig:benchmark_comp_TempDRL_VIC}. This considerable improvement has surpassed both the LP\&O and the TP\&O benchmarks (shown in the green dotted and red dash-dot lines, respectively) by approximately $23\%$ and $18\%$ in VIC, equivalent to AU\$$6,638$ and AU\$$5,366$ in total, respectively.

Also, it is noteworthy that, though the transformer model is more advanced than the LSTM model in time-series forecasting, its simulation results do not show significant superiority in better financial performance compared to the LP\&O method. The bidding outcomes of the LP\&O method even surpass that of the TP\&O method under multiple scenarios as shown in Table \ref{tab:benchmark_comp}, e.g., the contingency FCAS market of NSW. The results may suggest that even accurate forecast results may lead to poor decision-making, which is consistent with findings in recent research. The works by \cite{donti2017endtoend} and \cite{chen2023} attributed it to the uncoupled forecast model and optimization solver under the P\&O optimization framework. The former (i.e., the forecast model) is trained to improve the prediction accuracy rather than optimize the bidding decisions, making the forecast quality and the decision optimality decoupled, leading to likely unsatisfactory performance.

In addition, to evaluate the absolute performance of our TempDRL, we also implement an optimization benchmark with perfect future information on energy prices in the evaluation dataset (i.e., two months of prices). This benchmark is referred to as the \textbf{p}erfect \textbf{i}nformation \textbf{o}ptimization (PIO). Revenue comparisons of the TempDRL and the PIO method in the spot and contingency FCAS markets at five jurisdictions of the NEM are depicted in Fig. \ref{fig:benchmark_comp_tempDRL_PIO}, where the absolute bidding performance differences are also annotated in the percentage form. The absolute performance in the joint market scenario is shown in Fig. \ref{fig:benchmark_comp_tempDRL_PIO_joint} in Appendix B. The results reveal that our proposed TempDRL method achieves remarkable bidding outcomes compared to the PIO method with full knowledge of energy prices during the operational horizon of evaluation. In particular, the BESS's performance in the contingency FCAS markets of VIC and NSW nearly approximates the PIO method with gaps of only $16\%$. Moreover, as future information is unforeseeable in the electricity market, our proposed approach is more viable, practical, and applicable for the BESS's real-time bidding, which has better capability to mitigate market uncertainty and make profitable arbitrage decisions, compared to benchmarks, including the LP\&O, TP\&O, and MLP-DRL.

Furthermore, the TempDRL's performance with various sizes of the training dataset is also examined. Here, we re-train our TempDRL strategy using the one-month, three-month, and six-month training dataset, with evaluation revenues in the spot and contingency FCAS markets of the VIC in Fig. \ref{fig:train_size_comp_TempDRL_VIC}. The result of the joint-market bidding is presented in Fig. \ref{fig:train_size_comp_TempDRL_VIC_Join} in Appendix C. Such bidding outcomes suggest that an adequate amount of training data is essential to train a well-performed DRL-based bidding strategy, since its bidding performance is significantly improved with more available training data.

\subsubsection{Running Time Comparisons}
We evaluate the running time costs of the LP\&O and the TP\&O benchmarks, along with DRL-based methods, to examine their viability in real-time bidding. The results are shown in Table \ref{tab:running_time}. Specifically, we divide the algorithms' running time into the training and evaluation stages. The training stage includes model training time for the price predictors (in LP\&O and TP\&O benchmarks) and DRL-based strategies (i.e., MLP-DRL and TempDRL). The evaluation stage calculates the time costs of real-time bidding decisions in the two-month-length evaluation dataset. As shown in Table \ref{tab:running_time}, despite the longer training time of both the MLP-DRL and the TempDRL, our DRL-based bidding strategies can make significantly faster bidding decisions in the evaluation stage, i.e., $40$ seconds in total for two-month bidding. In this regard, benefiting from massive historical energy price data, a well-trained TempDRL strategy is better suited to real-time online bidding, where accurate and rapid decision-making is in particular crucial. Though training a DRL-based bidding strategy with feature extraction takes approximately $12$ more minutes, there is not a significant difference in the time consumed for bidding using MLP-DRL and TempDRL (with TTFE) during evaluation, spending $0.5$ and $0.7$ minutes, respectively.

\begin{table}[!t]
    \centering
    \caption{Running time of our TempDRL with/without the TTFE, the LP\&O benchmark, and the TP\&O benchmark in the joint-market bidding.}
    \begin{tabular}{c|c|c|c}
    
    \hline
    
    Method & Training Time & Evaluation Time & Total Time Cost \\
    
    \hline
    
    LP\&O & $3.8$ mins & $42.7$ mins & $46.5$ mins \\
    
    \hline 

    TP\&O & $5.2$ mins & $43.9$ mins & $49.1$ mins \\ 
    
    \hline
    
    MLP-DRL & $68.2$ mins & $\bm{0.5}$ mins & $68.5$ mins \\
    
    \hline
    
    TempDRL & $80.7$ mins & $\bm{0.7}$ mins & $80.5$ mins \\ 
    
    \hline
    \end{tabular}
    \label{tab:running_time}
\end{table}

\subsection{BESS Bidding Behavior Analysis} \label{subsec:exp_strategy_analysis}
To evaluate the bidding behavior of our proposed TempDRL in response to stochastic price signals, we conduct the analysis of the bidding decisions generated by the BESS in both individual and joint markets.

\begin{figure}[!t]
    \centering
    \subfloat[Charge.]{
    \includegraphics[width=.47\linewidth]{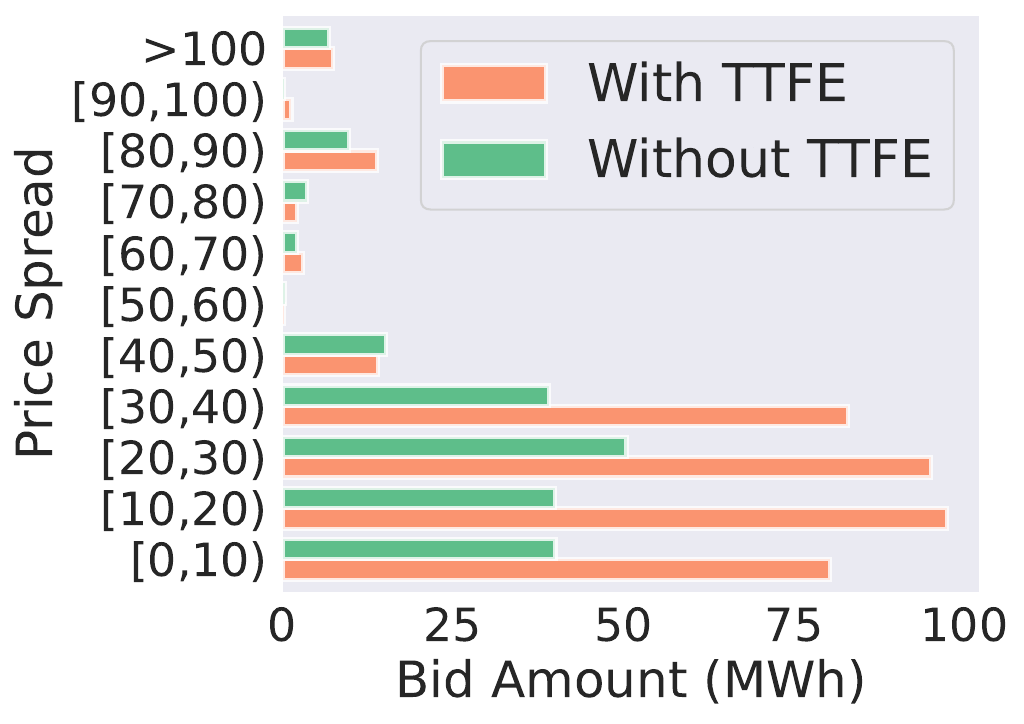}
    \label{fig:price_spread_spot_market_ch}
    }
    \subfloat[Discharge.]{
    \includegraphics[width=.47\linewidth]{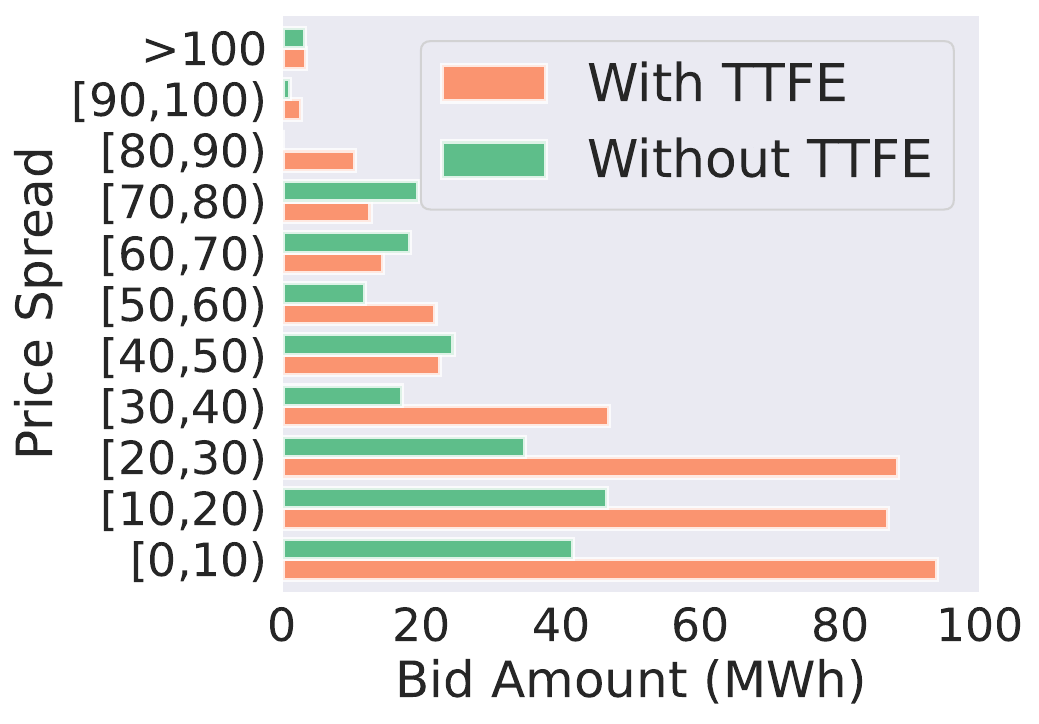}
    \label{fig:price_spread_spot_market_dch}
    }
    \caption{Bid amounts for arbitrage under different price spreads.}
    \label{fig:price_spread_spot_market}
\end{figure}

\begin{figure}[!t]
    \centering
    \includegraphics[width=0.8\linewidth]{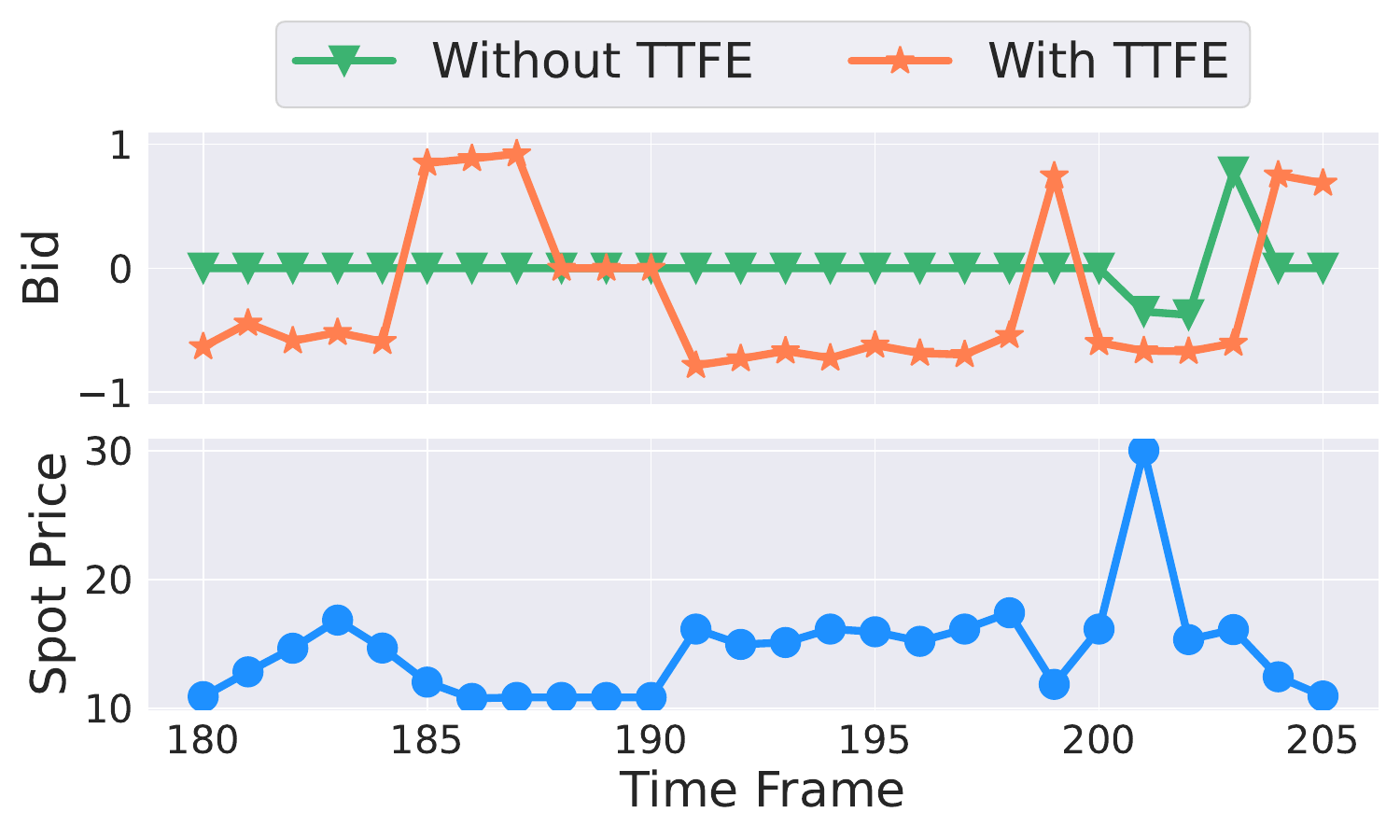}
    \caption{Energy arbitrage at below-average spot prices. Negative/positive values in the ``Bid'' axis indicate discharge/charge bids.}
    \label{fig:arbitrage_exmaple}
\end{figure}

\subsubsection{Spot Market} \label{subsubsec:exp_strategy_analysis_spot}
In the spot market, the BESS aims to take advantage of price spreads for energy arbitrage, i.e., buy low and sell high. In Fig. \ref{fig:price_spread_spot_market}, we illustrate the BESS's arbitrage behaviors via its bid amount (in MWh) under different price spreads when conducting arbitrage operations (in which charging and discharging decisions are shown in Fig. \ref{fig:price_spread_spot_market_ch} and \ref{fig:price_spread_spot_market_dch}, respectively). The results reveal that the DRL-based strategies with/without the TTFE perform similarly when there are significantly large price fluctuations, e.g., the price spread is larger than AU\$$40$/MWh under both charging and discharging scenarios. However, our TTFE-embedded strategy presents substantially better capability in utilizing relatively small price spreads to conduct more frequent arbitrage operations. The bid amounts of the TTFE-embedded strategy in price spread intervals, such as $[0,10)$, $[10,20)$, $[20,30)$, and $[30,40)$, are approximately twice more than those of the strategy without TTFE, as shown in Fig. \ref{fig:price_spread_spot_market}. Such a performance gap may be attributed to the introduction of our devised TTFE, which effectively captures temporal trends of energy prices and enables the BESS to be more responsive to small price spreads, reflected by significant revenue boosts in Table \ref{tab:benchmark_comp}.

In Fig. \ref{fig:arbitrage_exmaple}, the BESS can conduct energy arbitrage after identifying slight price fluctuations when equipped with our TTFE-embedded strategy, whereas the strategy without TTFE does not respond to relatively small price spreads during intervals $[180,200]$. These results highlight the advantage of incorporating temporal information in energy arbitrage decision-making. The interpretability of the attention mechanism used in our TTFE-embedded strategy is further discussed in Section \ref{subsubsec:interpretability_attn}.

\begin{table}[!t]
    \centering
    \caption{The evaluation results of the contingency FCAS market.}
    \begin{tabular}{c|c|c|c|c|c}
    \hline
    \multirow{2}{*}{\makecell{Sub \\ Market}} & \multirow{2}{*}{\makecell{Average \\ Price}}& \multicolumn{2}{c|}{\makecell{Contingency \\ Response Times}} & \multicolumn{2}{c}{Bid Power (MW)}\\
    \cline{3-6}
    & & No TTFE & TTFE & No TTFE & TTFE\\
    \cline{1-6}
    \centering FR & AU\$$5.44$ & \multirow{3}{*}{$236/341$} & \multirow{3}{*}{$\bm{302}/341$} & $6482$ & $\bm{9100}$ \\
    \cline{1-2}\cline{5-6}
    \centering SR & AU\$$3.25$ & & & $6692$ & $4094$ \\
    \cline{1-2}\cline{5-6}
    \centering DR & AU\$$2.93$ & & & $1625$ & $\bm{7576}$ \\
    \hline\hline
    \centering FL & AU\$$0.02$ & \multirow{3}{*}{$47/294$} & \multirow{3}{*}{$\bm{263}/294$} & $1253$ & $871$ \\
    \cline{1-2}\cline{5-6}
    \centering SL & AU\$$0.08$ & & & $1488$ & $928$ \\
    \cline{1-2}\cline{5-6}
    \centering DL & AU\$$0.50$ & & & $1581$ & $790$ \\
    \cline{1-6}
    \end{tabular}
    \label{tab:results_FCAS}
\end{table}

\begin{figure}[!t]
    \centering
    \includegraphics[width=0.8\linewidth]{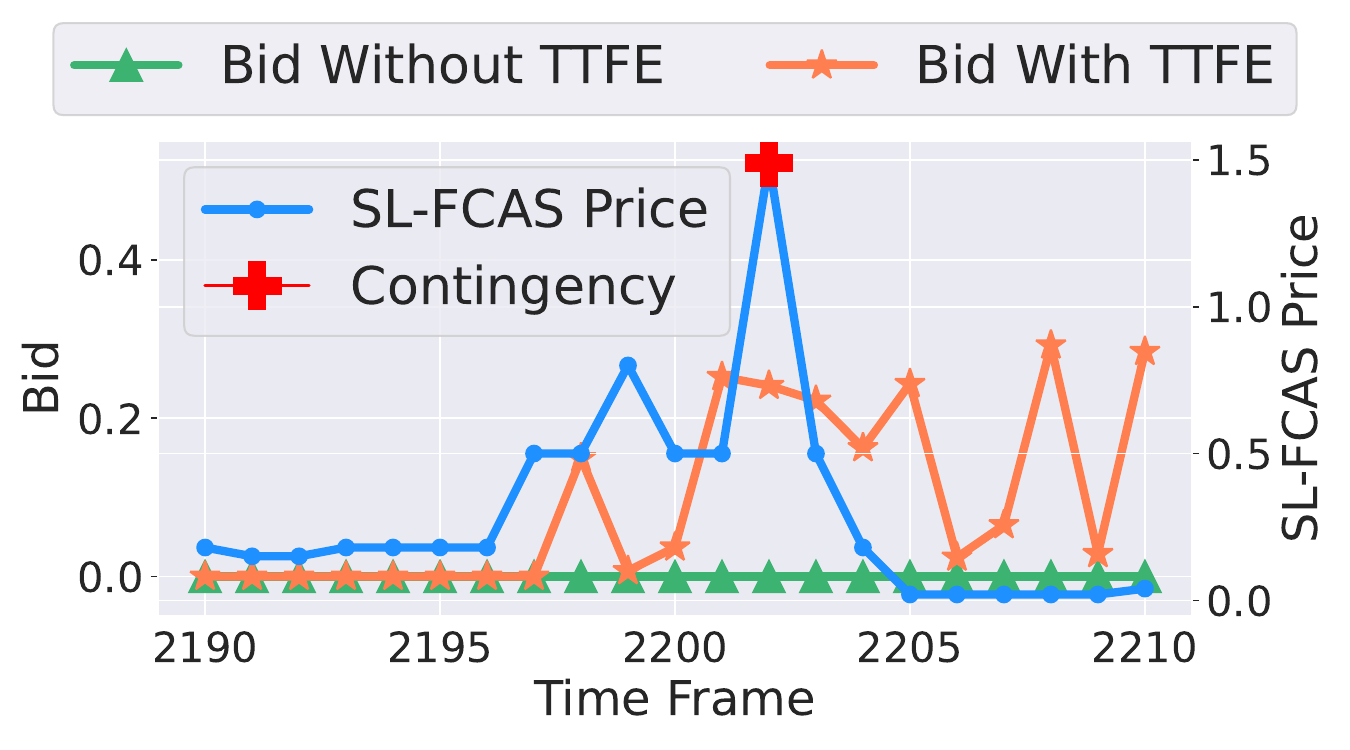}
    \caption{The bidding behaviors of the BESS in the SL sub-market.}
    \label{fig:results_SL-FCAS}
\end{figure}

\subsubsection{Contingency FCAS Market} \label{subsubsec:exp_strategy_analysis_FCAS}
Our proposed TTFE-embedded strategy also performs significantly better in the contingency FCAS market, where price spikes often occur during contingency events. The integration of temporal information in our strategy enables the BESS to be more responsive to price spikes in FCAS markets during its multi-FCAS-market participation, leading to improved performance compared to the strategy without the TTFE. This superiority is evident from the comparison results presented in Table \ref{tab:results_FCAS}.

The strategies with and without TTFE exhibit a significant difference in their response to major increases in system frequency, particularly in the provision of frequency-lower services in the FL, SL, and DL sub-markets. The strategy without TTFE is driven by the higher energy prices in the FR, SR, and DR sub-markets, leading to frequent bid for delivering frequency-raise services to increase profits. This behavior results in less focus on the frequency-lower sub-markets, which have relatively lower energy prices. However, providing frequency-lower services at contingency is the sole power source of the BESS in the contingency FCAS bidding scenario, and without adequate energy uptake from the grid, the BESS's SoC is getting drained and unable to continuously bid in the frequency-raise sub-markets due to a lack of available capacity.

In contrast, the TTFE-embedded strategy demonstrates superior ability in providing frequency-lower services at contingency, as shown in Table \ref{tab:results_FCAS}. Fig. \ref{fig:results_SL-FCAS} illustrates that the BESS, with the TTFE-embedded strategy, bids for partial power in the SL sub-market under varying energy prices and successfully delivers SL-FCAS during contingency, while the pure DRL-based strategy does not participate in the SL sub-market. The TTFE-embedded strategy allows the BESS to be more responsive to contingency events, particularly major rises in system frequency, as depicted in Fig. \ref{fig:results_SL-FCAS}. As a result, the BESS with ample energy storage is able to submit bids in the frequency-raise sub-markets more frequently, as shown in Table \ref{tab:results_FCAS}, thereby enhancing its economic performance.

\subsubsection{Joint Market} \label{subsubsec:exp_strategy_analysis_joint}
The BESS can fully realize its flexibility by participating in both the spot and contingency FCAS markets. The contingency FCAS market serves as an alternative to the spot market when prices in any of the six sub-markets are promising.

As mentioned, the use of extracted temporal information in the spot market greatly improves the BESS's viability, especially under small price fluctuations, but it still experiences idle periods of around $20\%$ of the evaluation time frame when it is completely drained or fully charged. This occurs when spot prices are exceptionally low or high, as highlighted in Table \ref{tab:idle_times_joint}. Such inaction reduces the BESS's revenue generation potential. Joint bidding in the contingency FCAS market eliminates such idle behavior and significantly improves the BESS's bidding outcomes, surpassing the results of individual market participation by a significant margin in all five jurisdictions of the NEM, as demonstrated in Table \ref{tab:benchmark_comp}.

\begin{table}[!t]
    \centering
    \caption{The BESS idle statistics in the spot and joint-market bidding.}
    \begin{tabular}{|c|c|c|c|c|}
    \hline
    \multirow{2}{*}{SoC Level}& \multicolumn{2}{c|}{Spot Price (AU\$/MWh)} & \multicolumn{2}{c|}{Idle Times} \\
    \cline{2-5}
    & Mean & Standard Deviation & Spot & Joint \\
    \cline{1-5}
    Empty & $3$ & $44$ & $3161$ & $\bm{0}$\\
    \cline{1-5}
    Full & $214$ & $47$ & $352$ & $\bm{0}$\\
    \cline{1-5}
    \end{tabular}
    \label{tab:idle_times_joint}
\end{table}

\begin{figure}[!t]
    \centering
    \includegraphics[width=0.9\linewidth]{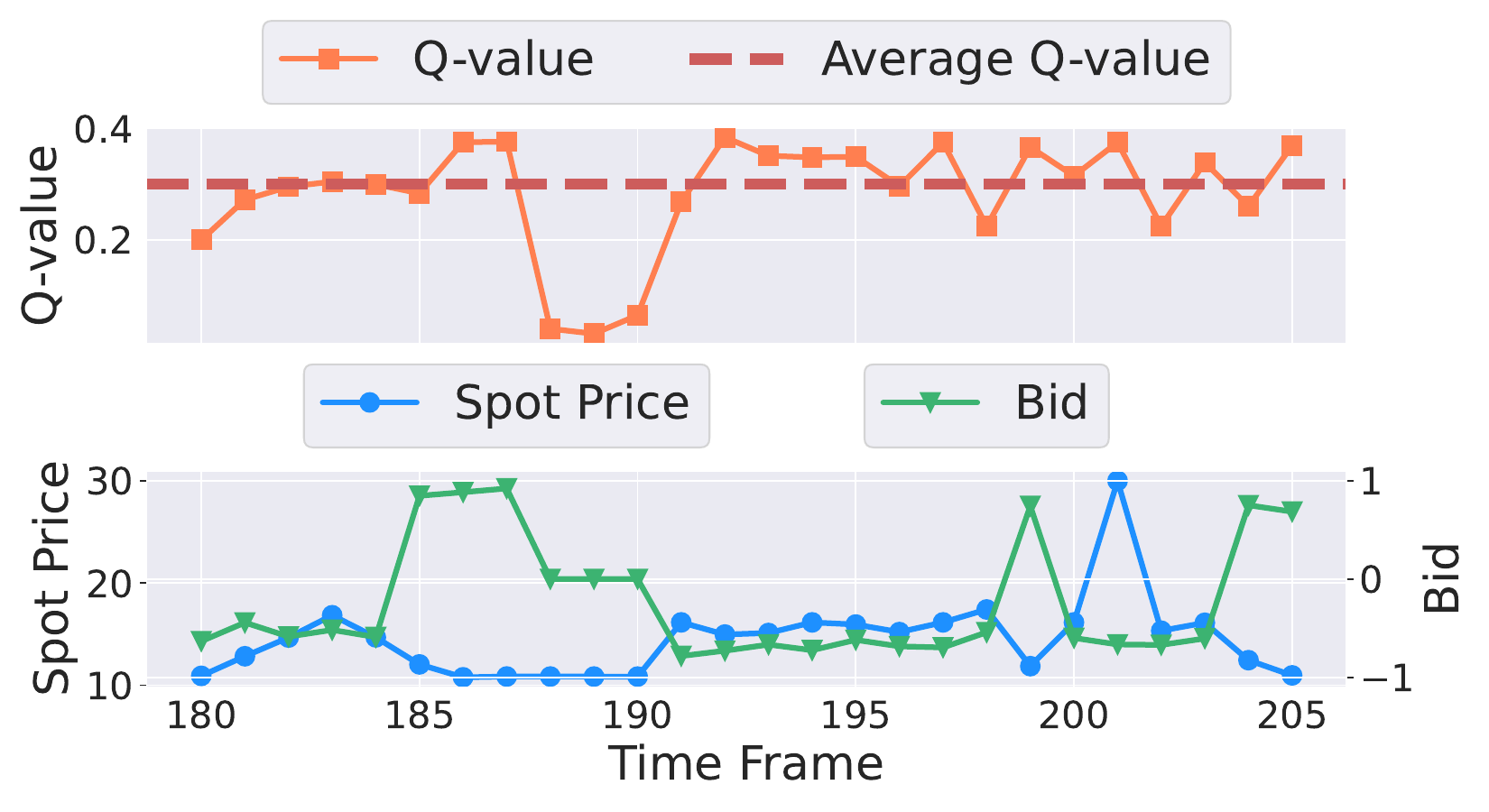}
    \caption{The normalized Q values when the BESS performs arbitrage.}
    \label{fig:q_val}
\end{figure}

\subsection{Interpretability of the TempDRL} \label{subsec:interpretability}
In DRL-based bidding strategies, there is an increasing need for a clear understanding of the internal decision-making mechanism, particularly how the model interprets the BESS's input state. To address this issue, we propose three new interpretations of our TempDRL model, utilizing Q-value, attention-based, and gradient-based approaches.

\subsubsection{Q-value-based Interpretation} \label{subsubsec:interpretability_qval}
The Q-value, represented by the function $Q(\bm{s}_t,\bm{a}_t)$, measures the effectiveness of a bidding decision based on the current state of the BESS. This value serves as a crucial indicator for evaluating the BESS's understanding and utilization of energy prices. In Fig. \ref{fig:q_val}, we display normalized Q-values when the BESS carries out energy arbitrage in the spot market (the same periods depicted in Fig. \ref{fig:arbitrage_exmaple}). Fig. \ref{fig:q_val} reveals that Q-values (when the BESS executes energy arbitrage) are significantly higher than those during idle periods, which are from the $188$th to $190$th time frames. This result suggests that, in our revenue-driven bidding strategy, idleness is discouraged and results in lower Q-values. This is because frequent arbitrage leads to a higher likelihood of generating greater revenue, even when encountering small price fluctuations, resulting in higher Q-values.

\begin{figure}[!t]
    \centering
    \includegraphics[width=0.8\linewidth]{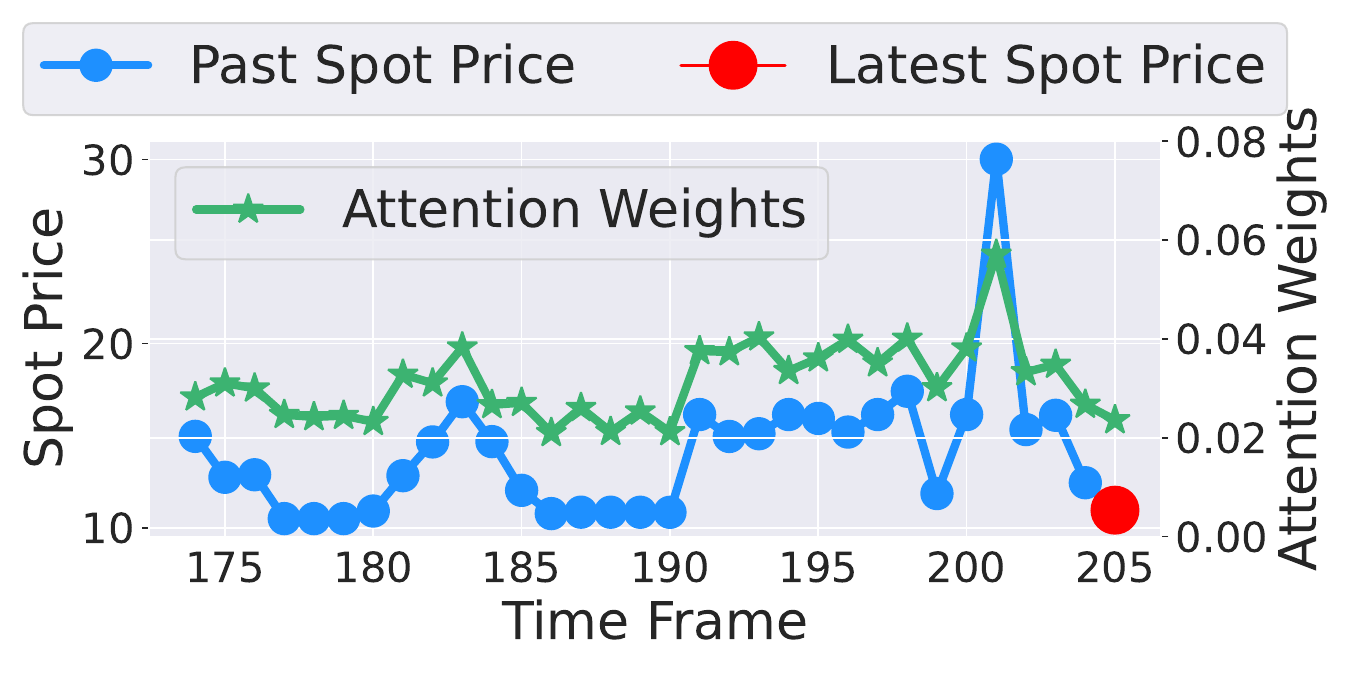}
    \caption{The relationship between attention weights and spot prices. The sum of attention weights is equal to one due to the softmax operation.}
    \label{fig:attn_weights}
\end{figure}

\subsubsection{Attention-based Interpretation} \label{subsubsec:interpretability_attn} 
The MHA mechanism functions as the key component of the developed TTFE and determines the relative temporal significance of each price vector based on how it is influenced by other price vectors. This allows the BESS to be aware of critical price signals. The attention weights of a temporal segment can be seen in Fig. \ref{fig:attn_weights}, where historical spot prices close to the latest spot price receive smaller attention weights, while those deviating from the latest spot price receive larger attention weights. This suggests that the TempDRL-empowered BESS is able to be temporally aware and sensitive to past price fluctuations when making bidding decisions. With the aid of the extracted temporal information, the BESS can better perform energy arbitrage in the spot market and be more responsive to contingency events in the contingency FCAS market, which is consistent with our findings in Section \ref{subsubsec:exp_strategy_analysis_spot} and \ref{subsubsec:exp_strategy_analysis_FCAS}.

\begin{figure}[!t]
    \centering
    \includegraphics[width=\linewidth]{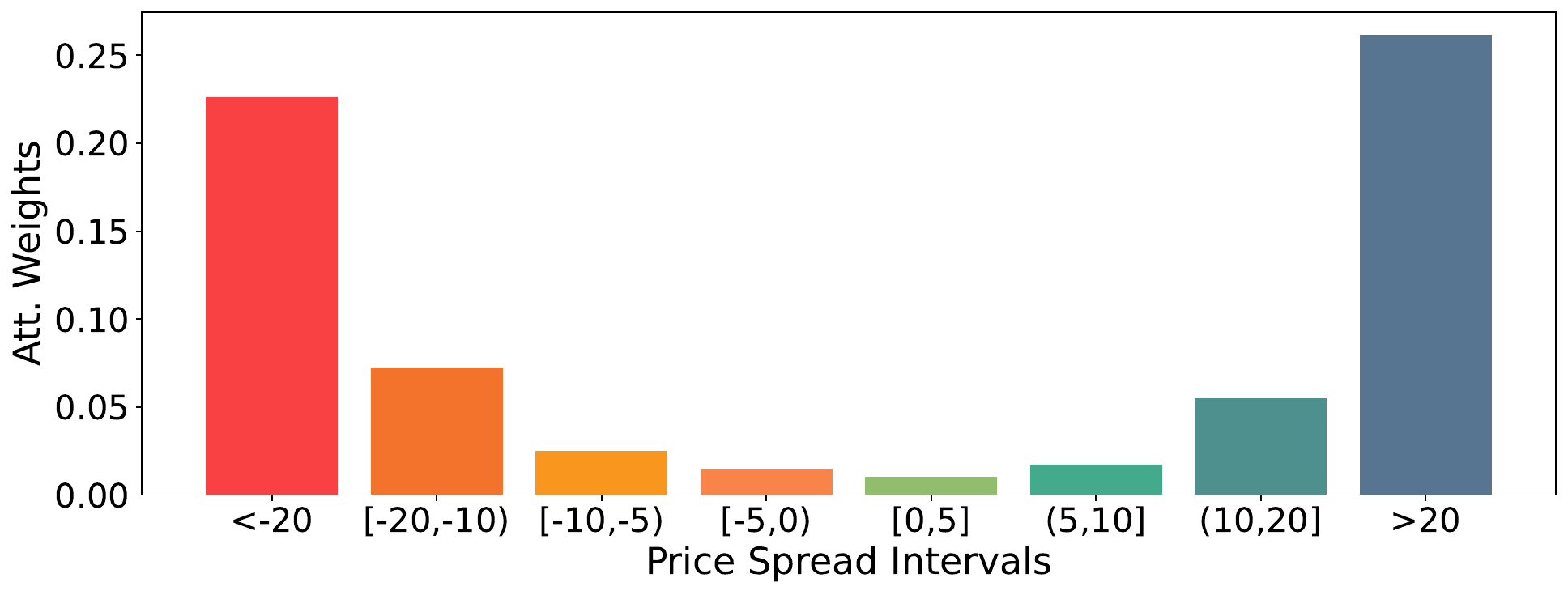}
    \caption{The distribution of price spreads with respect to attention weights.}
    \label{fig:price_spread_att_weight_dist_bar}
\end{figure}

To justify our implication, we analyze the relationship between \textit{the price spreads of latest and historical prices} and \textit{the corresponding attention weights of historical prices} in the whole test dataset. Specifically, for each input price sequence of length $L$, denoted as $\bm{\rho}_{t-1}=[\rho_{t-1-L}^\text{S},\cdots,\rho_{t-1}^\text{S}]$, we calculate the price spread between the latest spot price $\rho_{t-1}^\text{S}$ and each historical price $\rho_{t-1-L+t'}^\text{S}$, where $t'$ is an integer index variable from $0$ to $L-1$. A pair of the price spread, i.e., $\rho_{t-1}^\text{S}-\rho_{t-1-L+t'}^\text{S}$, and historical price's attention weight $w_{t-1-L+t'}^\text{att}$ is recorded and employed to derive the empirical distribution of price spreads to attention weights. The distribution is depicted in Fig. \ref{fig:price_spread_att_weight_dist_bar}, where we discretize the continuous price spread into eight intervals, including less than $-20$, $[-20,-10)$, $[-10,-5)$, $[-5,0)$, $[0,5]$, $(5,10]$, $(10,20]$, and more than $20$, and the attention weights within each interval are averaged. The results demonstrate that historical prices (which are \textit{lower or higher} than the latest spot price) gain larger attention weights, especially with the increase of price spreads, suggesting that our model tends to focus on fluctuating historical prices when making bidding decisions.

\begin{figure}[!t]
    \centering
    \includegraphics[width=\linewidth]{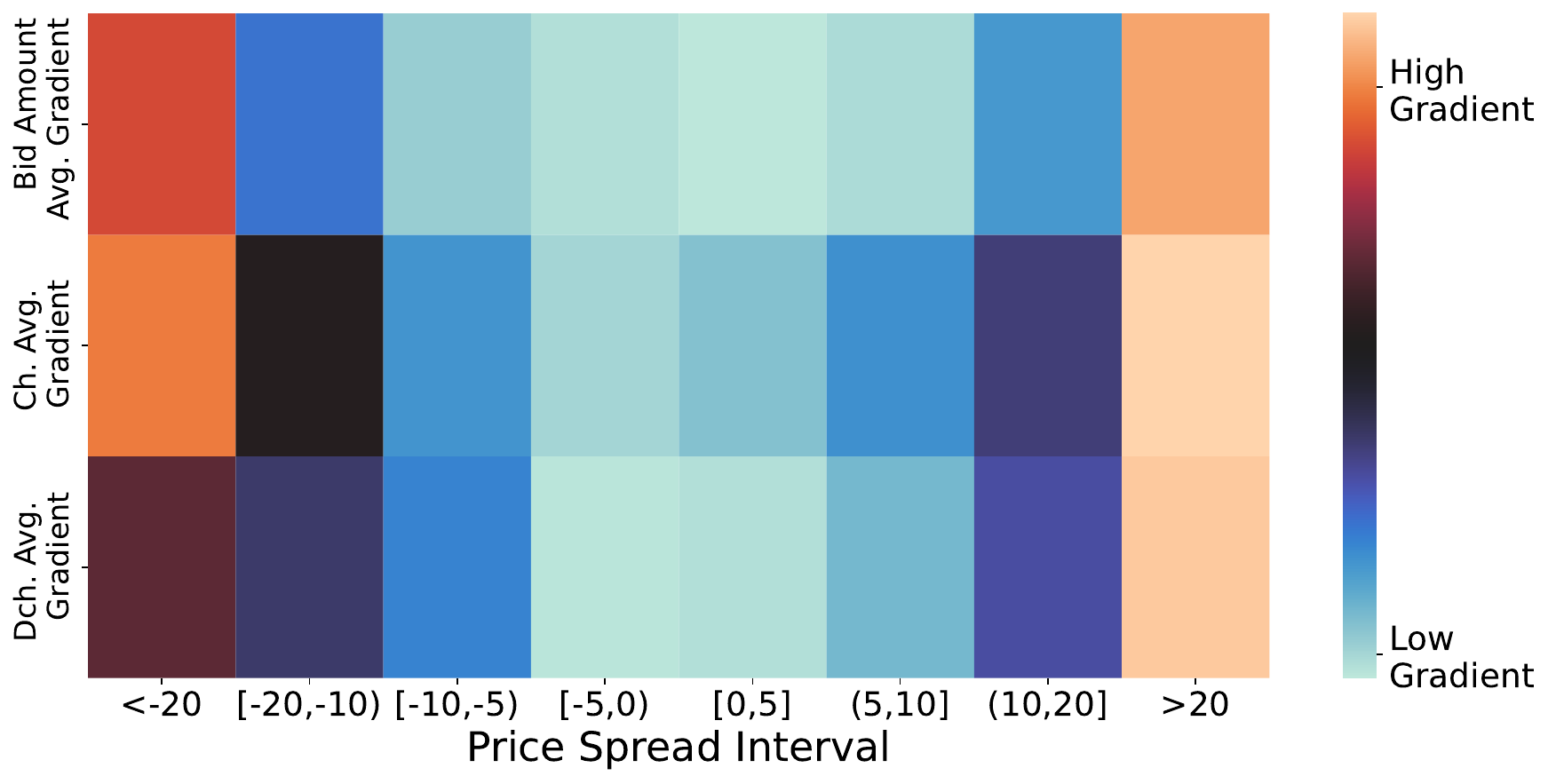}
    \caption{The gradient map of bidding decisions with respect to price spreads.}
    \label{fig:gradient_map}
\end{figure}

\subsubsection{Gradient-based Interpretation} \label{subsubsec:interpretability_grad}
In addition to the above Q-value-based and attention-based interpretations, we also investigate whether our model is capable of employing price fluctuation information via a gradient-based method. Specifically, we calculate the gradients of bidding decisions (including both charging/discharging and bid power) with respect to the price spreads (between latest price and historical prices). As the neural networks are updated by gradient backpropagation during the gradient descent process, the computed gradients can, to some extent, present the significance of how the input price sequence is related to the bidding outcome~\cite{avanti2017_gradients}. A higher value of the input price's gradient often indicates that this feature is emphasized and relatively more important during the decision-making process~\cite{sundararajan2017_gradients}.

The average gradients of charging decision, discharging decision, and bid power with respect to price spreads are depicted in Fig. \ref{fig:gradient_map}. In the gradient map, the gradients of price spreads, e.g., $[-20,-10)$ and $(10,20]$, are substantially larger than those of small price spreads, e.g., $[-5,0)$ and $[0,5]$. The results suggest that significant historical price fluctuations are indeed utilized, and impact final bidding decisions. Such findings are coherent with the attention-weight-based interpretations shown in Fig. \ref{fig:price_spread_att_weight_dist_bar}, demonstrating our model's ability to capture price fluctuations for better decision-making.

\section{Conclusion and Future Work}\label{sec:conclusions}
We developed a model-free, temporal-aware DRL strategy, referred to as TempDRL, for real-time joint bidding in the spot and contingency FCAS markets. Our proposed TTFE can capture temporal information of energy prices in both markets, enabling our TempDRL strategy to be sensitive to price fluctuations and make better bidding decisions. Our results showed that: 1) bidding in the joint markets can greatly enhance the viability of the BESS; and 2) the TTFE-empowered DRL-based bidding strategy can make better decisions, lead to significant outcomes compared to the P\&O benchmark. Additionally, our thorough analysis of bidding behaviors in both individual and joint markets revealed three insights: 1) the TTFE enhances the bidding strategy's ability to be aware of price fluctuations for optimal energy arbitrage in the spot market; 2) the TTFE-assisted bidding strategy is more responsive to contingency events in the contingency FCAS market; and 3) joint bidding leverages the flexibility of both markets to fully unlock the potential of the BESS. Moreover, the simulation results were interpreted based on the Q value and attention weights, providing a better understanding of the bidding decision-making mechanism of the TempDRL.

Our proposed strategy provides a methodological framework that the BESS owners can leverage to explore economic opportunities in other prospective markets. As part of the Post-2025 Market Design Options Paper \cite{ESB_reform}, the Energy Security Board (ESB) of Australia provides a vision of a future in which the system services necessary for securing and stabilizing the NEM are unbundled, allowing market participants to capitalize on alternative revenue streams, such as the provision of inertia, system strength and ramping capabilities, etc. A spot market-based approach is expected to become the norm eventually (as it is for energy and FCAS at the moment) for procuring those essential services \cite{ESB_reform}. This means that the owners of storage facilities, in particular utility-scale batteries, are able to generate additional revenues outside the traditional energy and FCAS markets. Our methodology may assist in harnessing these emerging opportunities. As Australia is at the forefront of the global energy transition, the successful demonstration of our strategy in the Australian electricity system could provide valuable insights for other countries embarking on the transition to a clean energy system.

For future work, we aim to study the influence of bidding strategies on grid emission, which is a critical metric for the net-zero transition of the grid. We will also investigate more realistic battery degradation characteristics of different types of energy storage in grid bidding applications.

\newpage
\section*{Appendix A: Revenue comparisons of the TempDRL and benchmarks in NSW, QLD, SA, and TAS} \label{appendix:revenue_comp}

\begin{figure}[!h]
    \centering
    \subfloat[NSW -- MLP-DRL]{
    \includegraphics[width=.48\linewidth]{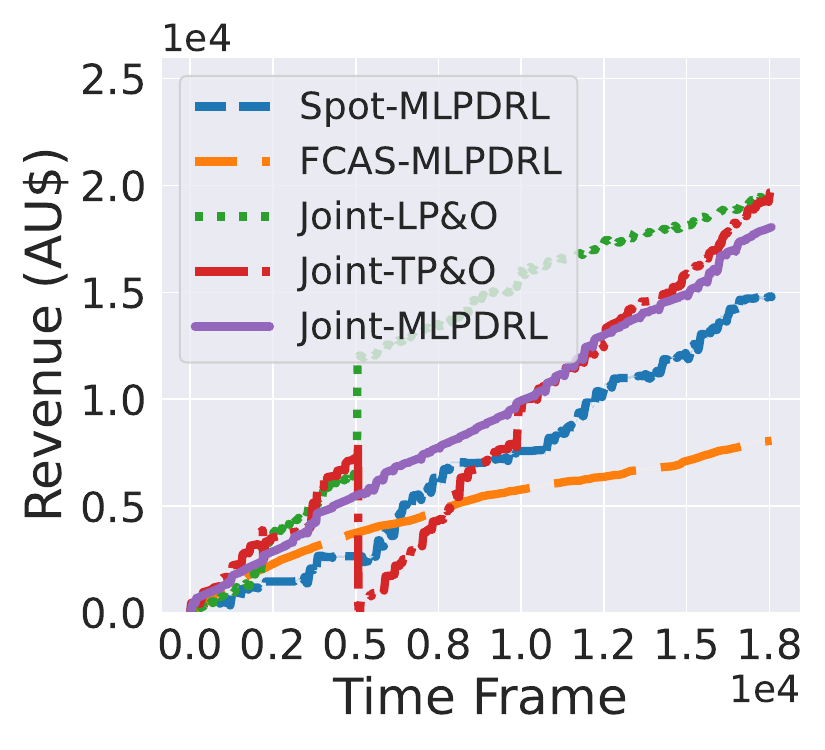}
    \label{fig:benchmark_comp_MLPDRL_NSW_R3C4}
    }
    \subfloat[NSW -- TempDRL]{
    \includegraphics[width=.48\linewidth]{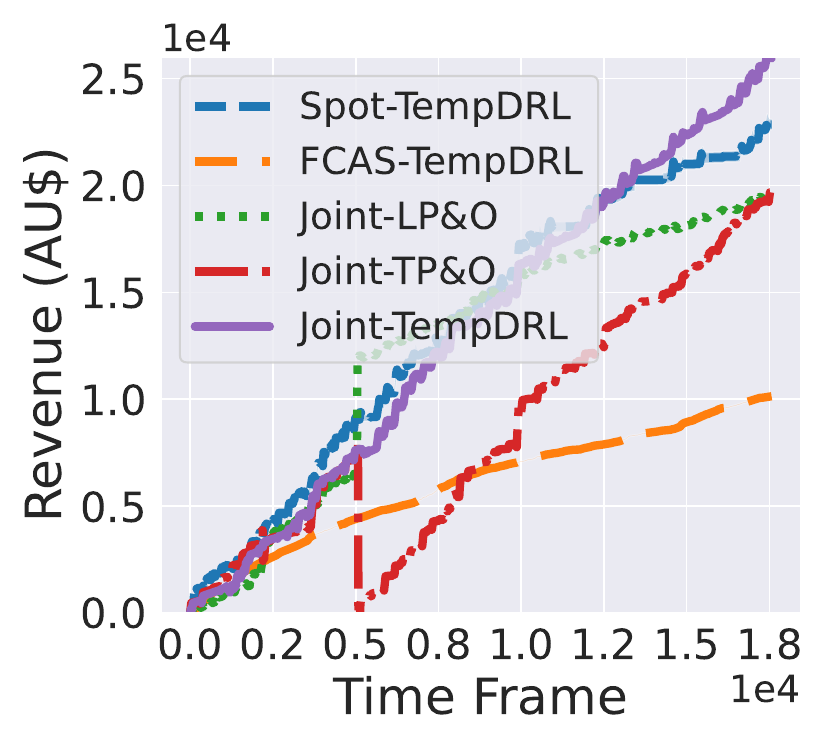}
    \label{fig:benchmark_comp_TempDRL_NSW_R3C4}
    }\\
    \subfloat[QLD -- MLP-DRL]{
    \includegraphics[width=.48\linewidth]{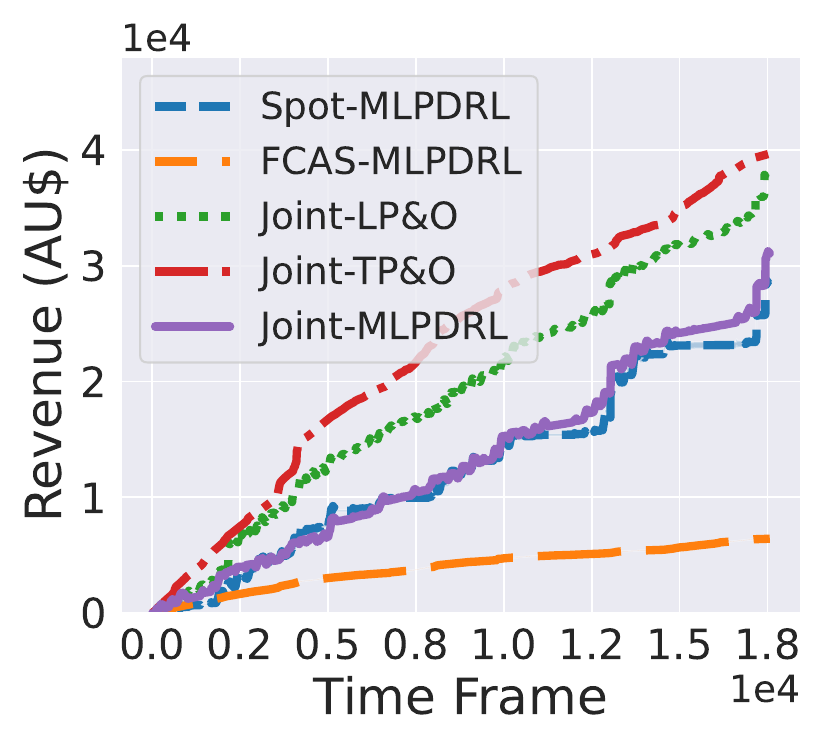}
    \label{fig:benchmark_comp_MLPDRL_QLD_R3C4}
    }
    \subfloat[QLD -- TempDRL]{
    \includegraphics[width=.48\linewidth]{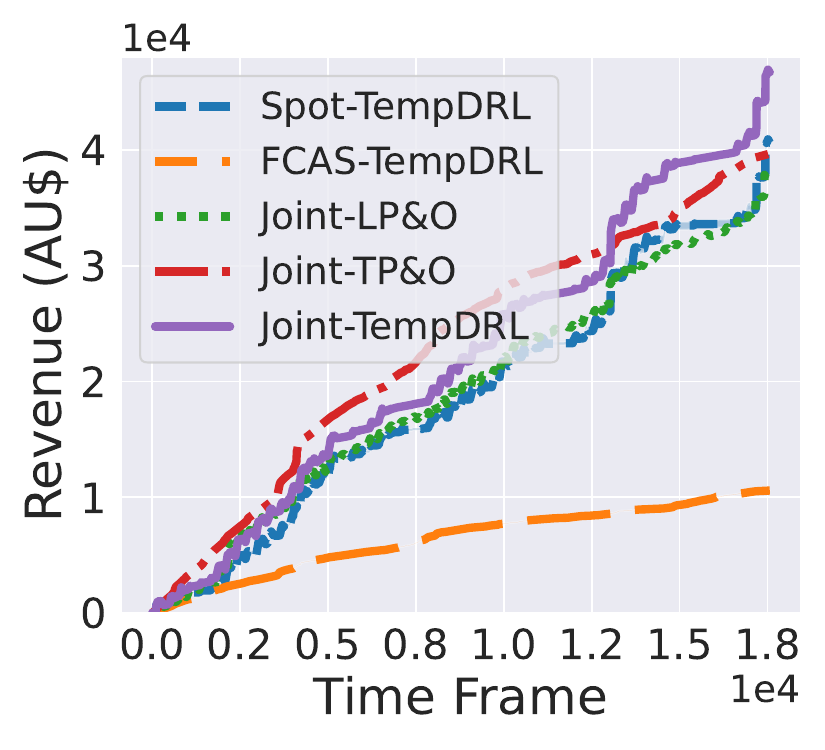}
    \label{fig:benchmark_comp_TempDRL_QLD_R3C4}
    }\\
    \subfloat[SA -- MLP-DRL]{
    \includegraphics[width=.48\linewidth]{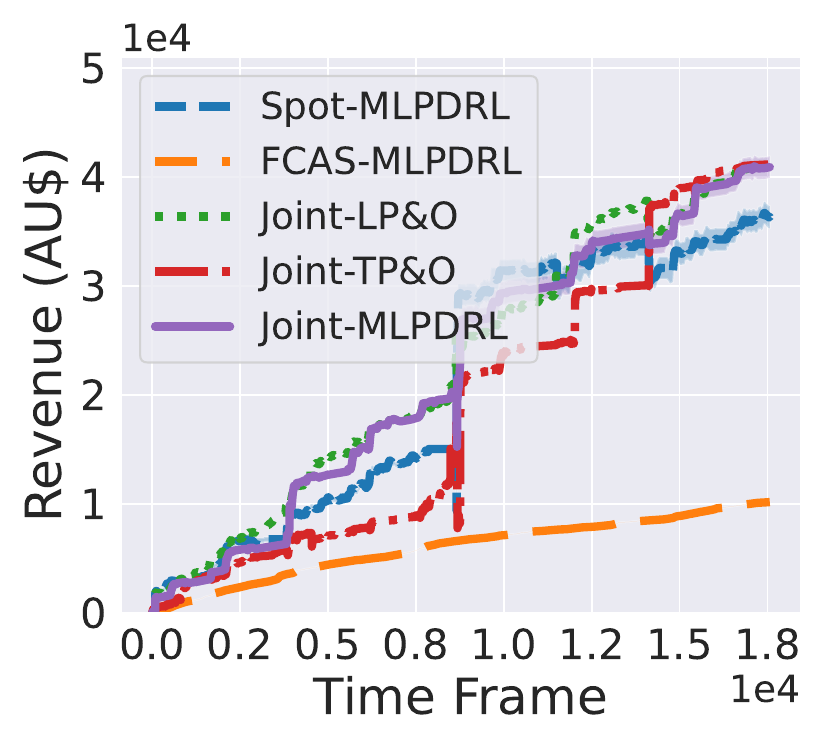}
    \label{fig:benchmark_comp_MLPDRL_SA_R3C4}
    }
    \subfloat[SA -- TempDRL]{
    \includegraphics[width=.48\linewidth]{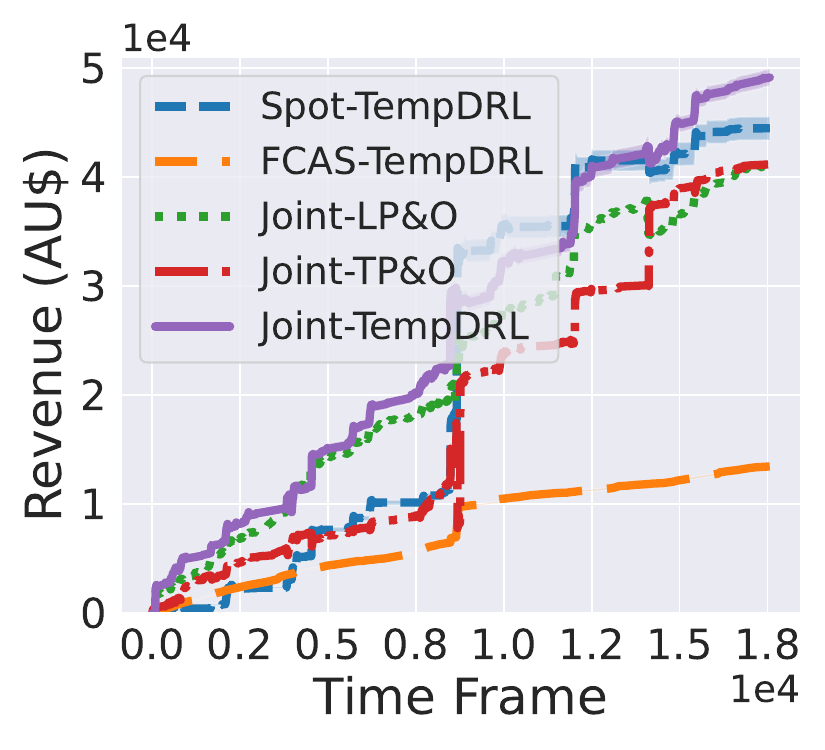}
    \label{fig:benchmark_comp_TempDRL_SA_R3C4}
    }\\
    \subfloat[TAS -- MLP-DRL]{
    \includegraphics[width=.48\linewidth]{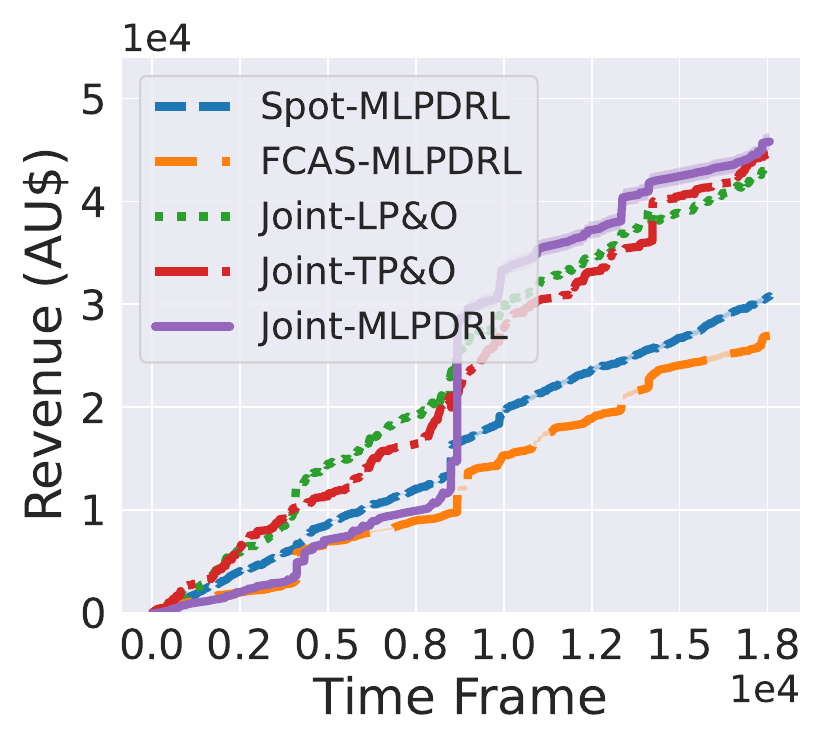}
    \label{fig:benchmark_comp_MLPDRL_TAS_R3C4}
    }
    \subfloat[TAS -- TempDRL]{
    \includegraphics[width=.48\linewidth]{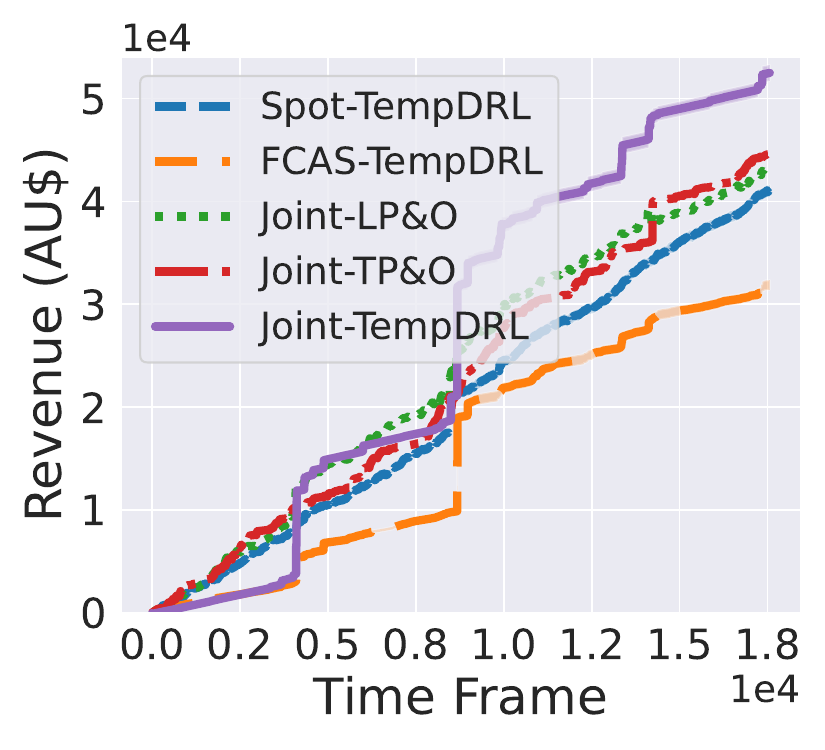}
    \label{fig:benchmark_comp_TempDRL_TAS_R3C4}
    }
    \caption{Revenue comparisons of the TempDRL method with benchmarks in spot, contingency FCAS, and joint markets of NSW, QLD, SA, and TAS.}
    \label{fig:benchmark_comp_NSW_QLD_SA_TAS}
\end{figure}

\newpage

\section*{Appendix B: Absolute Performance of the TempDRL and the PIO Benchmark in the Joint Market Scenario} \label{appendix:revenue_comp}

\begin{figure}[!h]
    \centering
    \includegraphics[width=.7\linewidth]{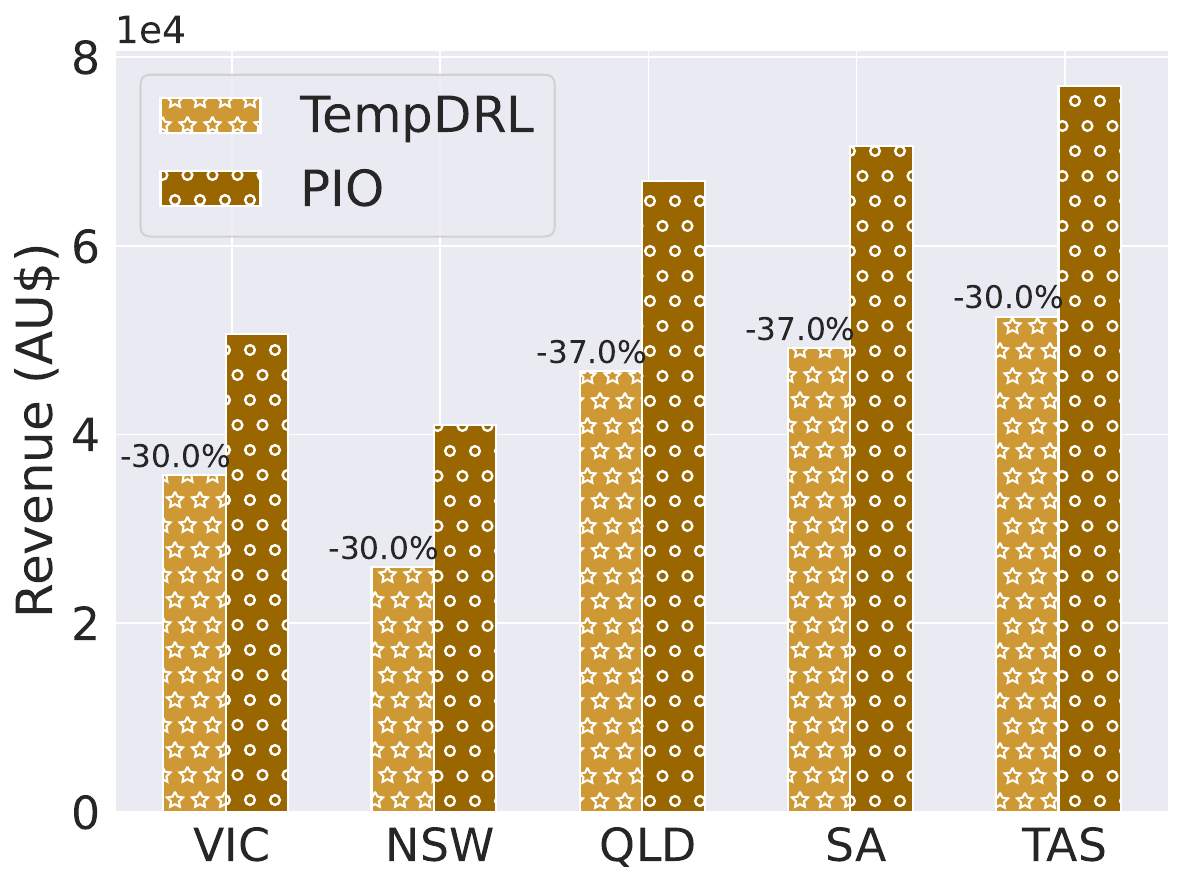}
    \caption{Revenue comparisons of the TempDRL and the PIO in the joint market at five jurisdictions of the NEM.}
    \label{fig:benchmark_comp_tempDRL_PIO_joint}
\end{figure}

\section*{Appendix C: Evaluation Revenues of the TempDRL Trained with Various Sizes of Dataset in the Joint-Market Bidding of VIC} \label{appendix:train_size_comp}

\begin{figure}[!h]
    \centering
    \includegraphics[width=.7\linewidth]{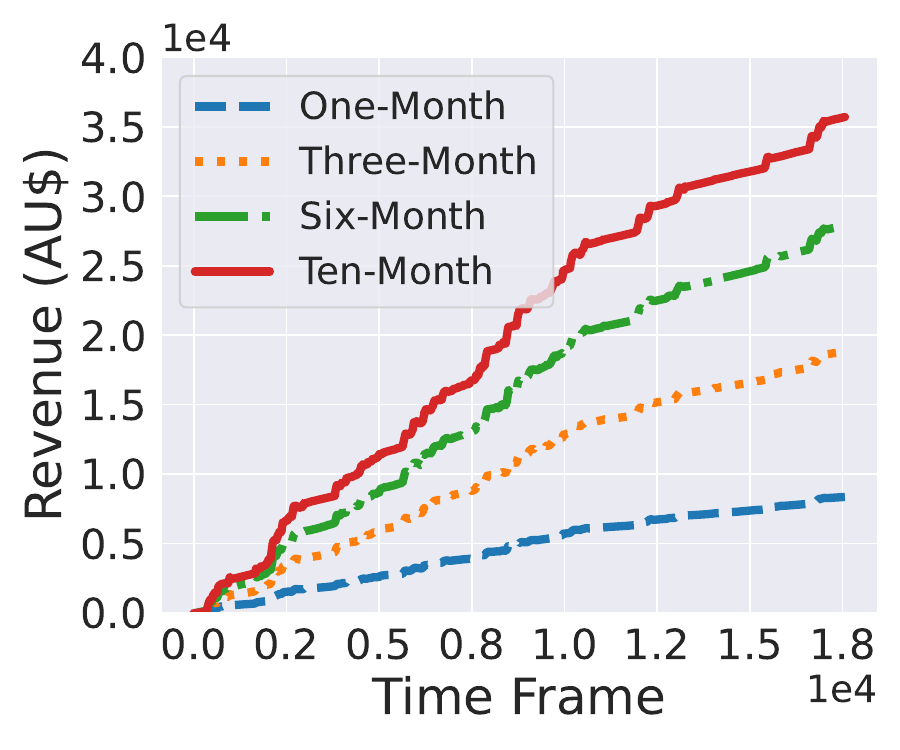}
    \caption{Revenue comparisons of the TempDRL trained with different sizes of dataset in the joint markets of VIC.}
    \label{fig:train_size_comp_TempDRL_VIC_Join}
\end{figure}

\bibliographystyle{IEEEtran}
\bibliography{IEEEabrv}


\begin{IEEEbiography}[{\includegraphics[width=1in,height=1.25in,clip,keepaspectratio]{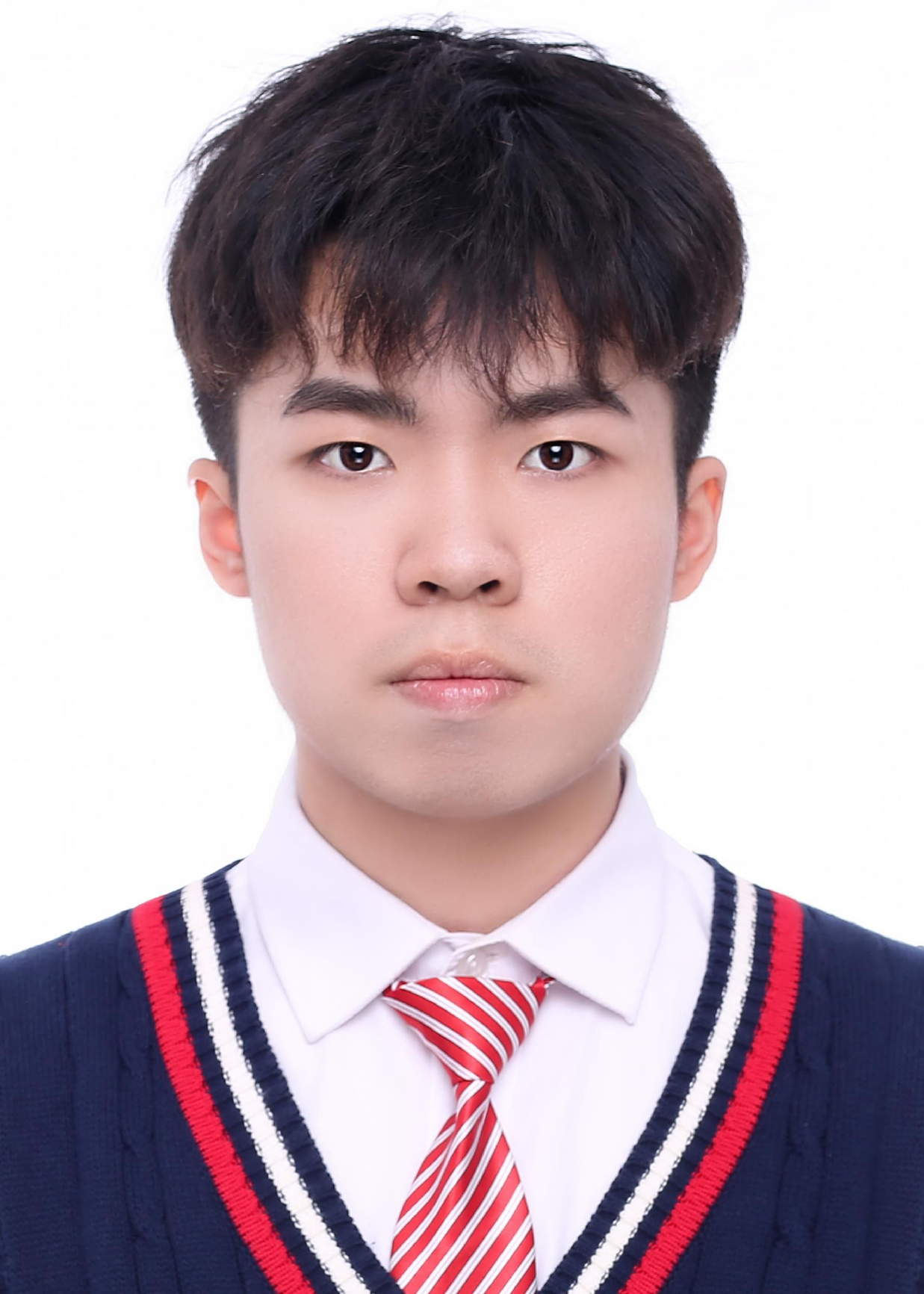}}]{Jinhao Li} received the B.E. degree in smart grid information engineering with a double B.S. degree in mathematics from the University of Electronic Science and Technology of China (UESTC), Chengdu, China, in 2022. He is currently working toward the Ph.D. degree with the Department of Data Science and AI, Faculty of Information Technology, Monash University, Melbourne, VIC, Australia. His research focuses on machine learning for energy systems.
\end{IEEEbiography}


\begin{IEEEbiography}[{\includegraphics[width=1in,height=1.25in,clip,keepaspectratio]{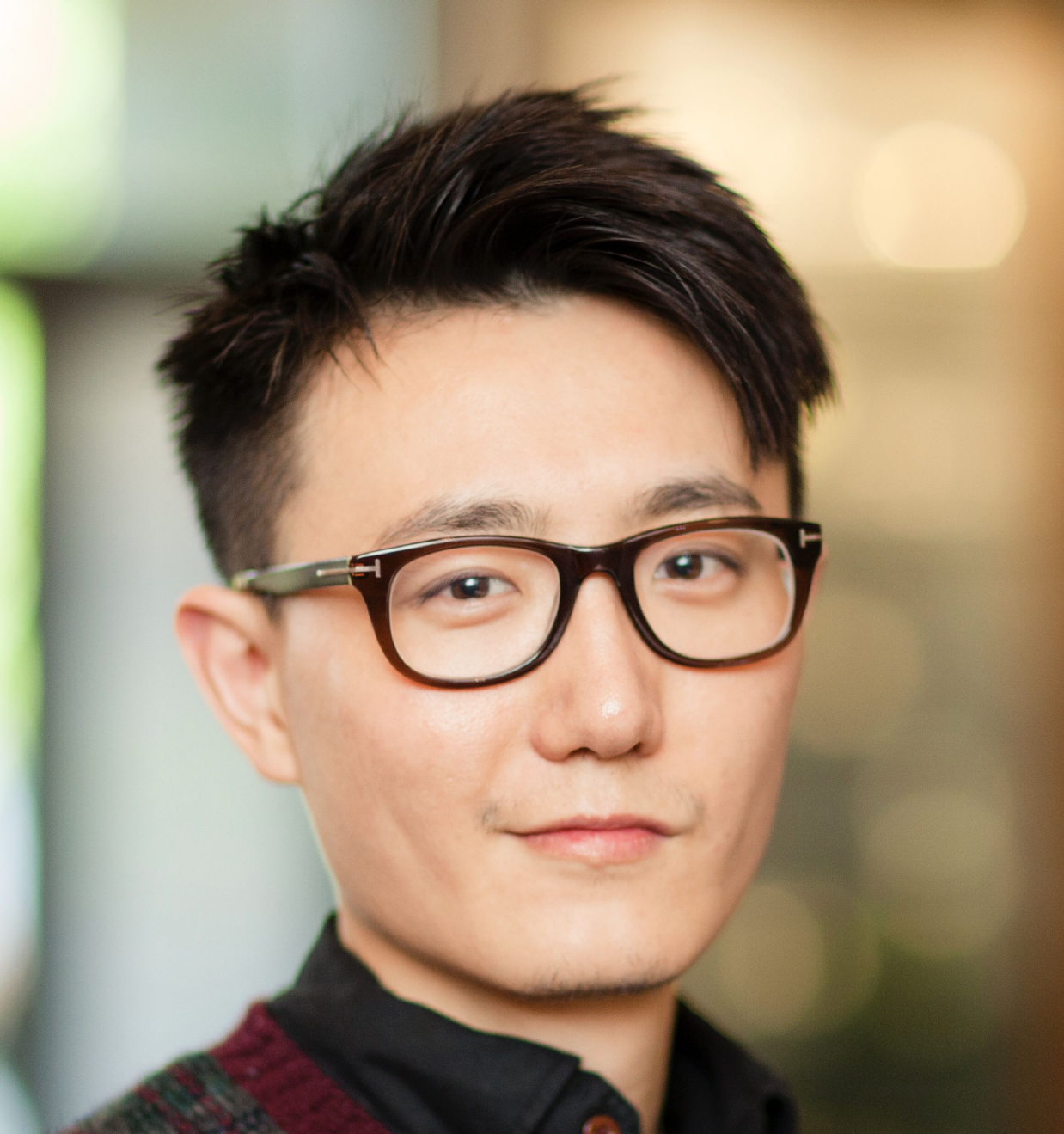}}]{Changlong Wang} received the Ph.D. degree in science from the University of Melbourne, Melbourne, VIC, Australia, in 2021. He is a Research Fellow with Monash University, Melbourne, VIC, Australia, specializing in energy system modeling. He is also a Climate Future Fellow with the University of Melbourne and a visiting scholar with the University of Oxford, Oxford, UK. He represents Australia on multiple IEA Hydrogen TCP tasks on hydrogen modeling. His Economic Fairways Mapper team was awarded the Australian Eureka Prize in 2023. 
\end{IEEEbiography}


\begin{IEEEbiography}[{\includegraphics[width=1in,height=1.25in,clip,keepaspectratio]{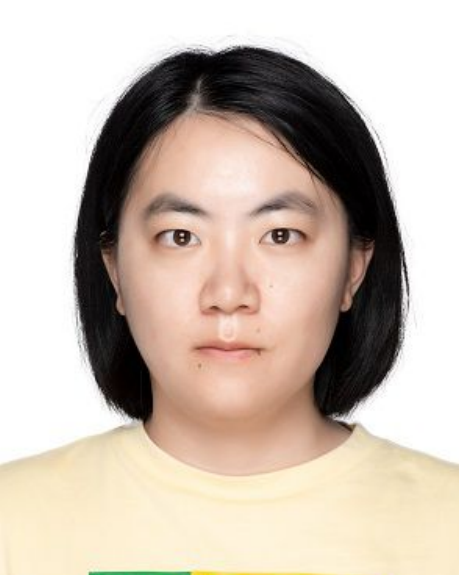}}]{Yanru Zhang} (S'13-M'16) received the B.S. degree in electronic engineering from the University of Electronic Science and Technology of China (UESTC), Chengdu, China, in 2012, and the Ph.D. degree from the Department of Electrical and Computer Engineering, University of Houston (UH), Houston, TX, USA, in 2016. She worked as the Postdoctoral Fellow at UH and the Chinese University of Hong Kong, Hong Kong, successively. She is currently a Professor with UESTC, affiliated with both Shenzhen Institute for Advanced Study and School of Computer Science and Engineering. Her research involves game theory, machine learning, deep learning in network economics, Internet and applications, wireless communications, and networking. She received the Best Paper Award with IEEE HPCC 2022, DependSys 2022, ICCC 2017, and ICCS 2016.
\end{IEEEbiography}

\begin{IEEEbiography}[{\includegraphics[width=1in,height=1.25in,clip,keepaspectratio]{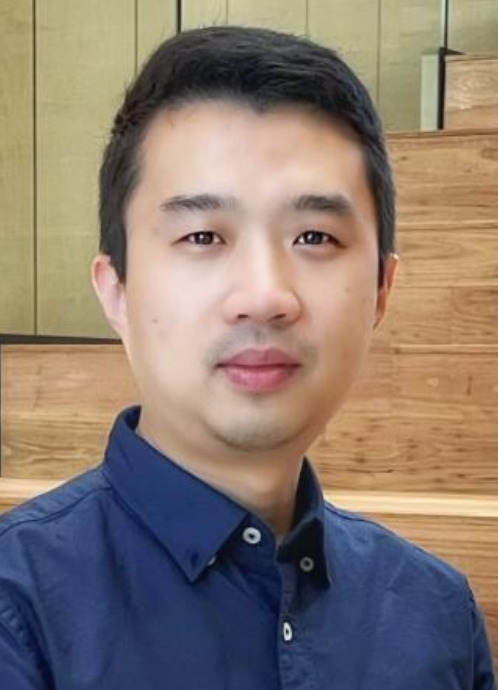}}]{Hao Wang} (M'16) received his Ph.D. in information engineering from The Chinese University of Hong Kong, Hong Kong, in 2016. He was a Postdoctoral Research Fellow at Stanford University, Stanford, CA, USA, and a Washington Research Foundation Innovation Fellow at the University of Washington, Seattle, WA, USA. He is currently a Senior Lecturer and ARC DECRA Fellow with the Department of Data Science and AI, Faculty of IT, Monash University, Melbourne, VIC, Australia. His research interests include optimization, machine learning, and data analytics for power and energy systems.
\end{IEEEbiography}

\end{document}